\documentclass[a4paper]{article}

\usepackage[margin=0.9in]{geometry} 
\usepackage{graphicx}
\usepackage{rotating}
\usepackage{amsmath}
\usepackage{mathrsfs}
\usepackage{amsfonts}
\usepackage{amssymb}
\usepackage{subfig}
\usepackage{xcolor}
\usepackage{float}
\usepackage{natbib}
\usepackage{setspace}

\usepackage{amsthm}

\usepackage[utf8]{inputenc}
\usepackage{hyperref}
\hypersetup{
	unicode,
	pdfauthor={Author One, Author Two, Author Three},
	pdftitle={Joint modelling of time-to-event and longitudinal response using robust skew normal-independent distributions},
	pdfsubject={Joint modelling of time-to-event and longitudinal response using robust skew normal-independent distributions},
	pdfkeywords={article, template, simple},
	pdfproducer={LaTeX},
	pdfcreator={pdflatex}
}

\theoremstyle{plain}

\theoremstyle{definition}

\usepackage{graphicx, color}
\graphicspath{{Figures/}}

\usepackage{algorithm, algpseudocode} 
\usepackage{mathrsfs} 

\usepackage{lipsum}

\title{Joint modelling of time-to-event and longitudinal response using robust skew normal-independent distributions}
\author{Srimanti Dutta$^1$\thanks{Address for correspondence: Division of Cancer Epidemiology, German Cancer Research Centre, Im Neuenheimer Feld 280, Heidelberg 69120, Germany; E-mail: \texttt{srimanti.dutta@dkfz-heidelberg.de}} \and Arindom Chakraborty$^2$ \and Dipankar Bandyopadhyay$^3$}

\date{
	$^1$Division of Cancer Epidemiology, German Cancer Research Centre, Heidelberg, Germany
             \\
	$^2$ Department of Statistics, Visva-Bharati University, Santiniketan, India \\
              $^3$ Department of Biostatistics, Virginia Commonwealth University, Richmond, VA, USA 
              \\
}

\begin{document}
\maketitle

\begin{abstract}
		Joint modelling of longitudinal observations and event times continues to remain a topic of considerable interest in biomedical research. For example, in HIV studies, the longitudinal bio-marker such as CD4 cell count in a patient's blood over follow up months is jointly modelled with the time to disease progression, death or dropout via a random intercept term mostly assumed to be Gaussian. However, longitudinal observations in these kinds of studies often exhibit non-Gaussian behavior (due to high degree of skewness), and parameter estimation is often compromised under violations of the Gaussian assumptions. In linear mixed-effects model assumptions, the distributional assumption for the subject-specific random-effects is taken as Gaussian which may not be true in many situations. Further, this assumption makes the model extremely sensitive to outlying observations. We address these issues in this work by devising a joint model which uses a robust distribution in a parametric setup along with a conditional distributional assumption that ensures dependency of two processes in case the subject-specific random effects is given.
		
 
\end{abstract}

\doublespacing

\section{Introduction} \label{sec:intro}
	
In many applied fields like epidemiology, pharmacokinetics etc. it has become increasingly common to record the values of a key longitudinal variable until the occurrence of possibly censored time-to-event (failure or survival) of a subject. For example, in studies of prostate cancer, repeated measurements of prostate-specific antigen (PSA) are collected for each patient following treatment, along with time to disease recurrence \citep{lin2002maximum}. In schizophrenia studies, longitudinal measurements are collected on the positive and negative syndrome scale (PANSS). Patients with higher PANSS scores indicate more severe psychopathy and have a higher probability of discontinuing the treatment because of low efficacy. Very often, interest focuses on an interrelationship between these variables. 

There are two linked sub-models in a joint model- one for a time-to-event and the other one describing the longitudinal process. Modelling these two simultaneous processes has received considerable attention over the past few decades. Starting from the work of Faucett and Thomas \citep{faucett1996simultaneously} and Wulfsohn and Tsiatis \citep{wulfsohn1997joint} the progression made in this direction is remarkable. Different factorizations of the likelihood result in different classes of joint models. In the literature, three broad classes of joint models have been discussed: selection models, pattern-mixture models and latent variable models \citep{xu2001joint}. A good review of different types of joint models can be found in work by Sousa \citep{sousa}. Both the pattern-mixture model \citep{little1993pattern} and the selection model \citep{little1989analysis}, \citep{little2000causal} describe exactly the same joint distribution, their statistical interpretations being different. A detailed overview of pattern-mixture and selection models can be found in the works of Fitzmaurice et al. \citep{fitzmaurice2008longitudinal} and Molenberghs et al. \citep{molenberghs2003sensitivity}.

In early literature on joint models, time-to-event data are modeled parametrically, which facilitates, in principle, straightforward likelihood inference (\citep{pawitan1993modeling}). Later on, proportional hazard models are used for time-to-event data. Tseng et al. \citep{tseng2005joint} adopted the accelerated failure time (AFT) model for the survival process in joint modelling. AFT models are predominantly fully parametric. The regression parameter estimates from AFT models are robust to omitted covariates.
 
 Mixed-effects models for repeated measures data have become popular because of their flexible covariance structure. Moreover, in mixed effect models it is assumed that individuals' responses follow a functional form (linear or nonlinear) with parameters varying among individuals. These models adopt normality assumptions for subject-specific random-effects that is highly sensitive to the presence of outliers and ineffective in the cases of departure from normality. 
 
  The traditional joint models available in the literature to date are restricted because when the subject-specific random-effects is given, the two processes become independent. This can be overcome by introducing a structural dependency between the two processes which means that even if the subject-specific random-effects are given the two processes will remain associated. This has been addressed by Dutta et al. \citep{dutta2021joint} where a novel model has been developed which captures the dependency by means of structural association. In that work, an $(n_i+1)$ dimensional random variable was introduced where the $(n_i+1)^{th}$ component represented the time-to-event observation which remained structurally associated with the $n_i$ variate longitudinal observation for the $i^{th}$ patient. This would ensure that the two processes were associated with each other via an extra layer of association apart from the association by means of subject-specific random-effects. For instance, in oncological studies, the time to death or worsening disease condition of a patient can be perceived as to some extent to be directly dependent on the tumour size of the patient.
  
   The association structure between the two sub models has always been vital when it comes to right interpretation of the model parameters. Exact identification of the underlying interrelationship between the two processes is essential as it leads to correct estimation of the treatment effects, reduced sample size in trials and increased power. Mchunu et. al \citep{mchunu2022using} had explored the clinical perspective in this context and had explored four alternative functional formas of association between the longitudianl and survival biomarkers. Further, Zou et. al \citep{zou2023bayesian} in their dynamic prediction framework for prediction of personalized disease progression had explored several functional forms of association between the two biomarkers.
  
  However, in longitudinal data analysis, many influential observations may occur. Wu \citep{wu2009mixed} classified outliers broadly into two categories i.e., one where there is the presence of outliers among repeated observations within an individual (individual not considered as an outlier) and another where an individual has been considered as an outlier owing to its differing behaviour from rest of the sample units.  
  
  A standard procedure to take care of these outliers is to consider some robust estimation techniques like M-estimation. Since the likelihood for a joint model is very complicated, M-estimation is likely to bring more complexity to the method. 
  
  To avoid this problem, we may consider a family of distributions that can take care of these influential observations to some extent.  These problems may be treated by adopting robust estimation techniques which uses heavy-tailed distributional assumptions in the mixed-effects models (\citep{lange1989robust}, \citep{rosa2003robust}). A multivariate t-linear mixed-model was proposed by Pinheiro et al. \citep{pinheiro2001efficient} and it was seen to perform efficiently when outliers were present in the data. In the context of joint modelling, Huang et al. \citep{huang2010joint} and Li et al. \citep{li2009robust} considered Student's t distribution having a fixed degree of freedom for robust modelling. Again, a heterogeneous linear mixed model was introduced by Verbeke and Lesaffre \citep{verbeke1996linear} where a finite mixture of normal formulated the distribution of subject-specific random-effects. Zhang and Davidian proposed a linear mixed model (LMM)  \citep{zhang2001linear} where random-effects were made to follow a semi-nonparametric distribution. 
  
  The skew-normal/independent distribution is another family of distributions that has been used to model the longitudinal response. In this context, a family of distributions is important where a standard normal variable ($Z$) is divided by an independent positive random variable ($V$) i.e., $Z/V$ where $V$ is independent of $Z$ (see \citep{andrews1974scale}). This $Z/V$ is called normal/independent (N/I) random variable which becomes a family of mixture distributions with heavy tails and hence becomes a suitable distribution for incorporating outliers.    

To estimate the effect of longitudinal outcomes on time-to-event, either a frequentist or Bayesian approach is used. The most common and well-understood frequentist approaches depend on the maximum-likelihood method. In many situations, this approach suffers from various computational issues like slow or non-convergence of the MCEM algorithm. To improve the analysis more flexible estimation method using related historical information can be achieved by employing a Bayesian approach (see \citep{chi2006joint}, \citep{wang2001jointly}, \citep{brown2003bayesian} and references therein). These approaches are generally based on Markov chain Monte Carlo (MCMC) sampling algorithms. Some versions of Gibbs sampling or Metropolis-Hastings (MH) sampler have been used for sample generation. To get faster convergence, instead of Gibbs sampler, Li and Luo considered No-U-Turn sampler[\citep{li2017}, \citep{liluo2019}]. Instead of an MCMC method, Hamiltonian Monte Carlo (HMC)  has been employed by Brilleman et al. \citep{brillman81}. This method explores the parameter space of the posterior distribution more efficiently. Wang et al. \citep{wang17} used a combination of HMC and No-U-Turn sampler.  Rosa et al. \citep{rosa2003robust} used Bayesian implementation of MCMC in the context of adopting normal/independent distribution under robust linear mixed-effect model. A similar type of work under a non-linear set up has been considered by Bandyopadhyay et al. \citep{bandyopadhyay2015robust}. Rustand et al. \citep{rustand2023fast} had introduced a Bayesian approximation based on the integrated nested Laplace approximation algorithm in R-INLA to deal with the computational burden in multivariate setup.

In this work, we deviate from the usual joint models in two ways. Firstly, we are using a more robust distribution in the parametric set up which can take care of influential observation to a great extent. Secondly, even if subject-specific random-effects is given in this model, the processes remain dependent. This is because the conditional distribution of the time-to-event variable $(T)$ given longitudinal variable $(Y)$ involves $Y$. This structural dependency can be considered more general than the traditional joint models where processes become independent if the subject-specific random-effects is given. However, if the correlation between two processes becomes zero, this model reduces to the traditional joint model. We have considered a Bayesian framework for inference. We have used OpenBugs and R for all computations.

The paper is organized as follows. In section 2, some relevant results for skew-independent distribution have been stated. Using the conditional distribution, a joint model has been introduced in Section 3. In this section, we have discussed a Bayesian approach in this situation. To judge the performance of the proposed methodology, detailed simulation studies have been performed. In each set-up, performance of all competing models with 5\% and 10\% outliers in covariates have been compared. In section 4, a well-known dataset on AIDS study has been analyzed. This paper concludes with some discussion.

\section{Multivariate Normal/ Independent (N/I) Distribution} \label{sec:model}
	
A skew normal independent or SNI distribution is a process of generating a p-dimensional random vector of the form
\begin{equation}
\boldsymbol{Y}=\boldsymbol{\mu}+U^{-1/2}\boldsymbol{W}
\end{equation}
where $\boldsymbol{\mu}$ is a location vector, $U$ is a positive random variable with cumulative distribution function $H(u|\nu)$ and probability density function $h(u|\nu)$, $\nu$ is a scalar or vector of parameters indexing the distribution of $U$, which is a positive value and $\boldsymbol{W}$ is a multivariate skew-normal random vector with location vector $\boldsymbol{0}$, scale matrix $\Sigma$ and skewness parameter vector $\lambda$ and we the notation $\boldsymbol{W} \sim SN_{p}(\boldsymbol{0} , \boldsymbol{\Sigma}, \boldsymbol{\lambda})$.   \\
Given $\boldsymbol{U}$, $\boldsymbol{Y}$ follows a multivariate skew-normal distribution with location vector $\boldsymbol{0}$, scale matrix $u^{-1}\Sigma$ and skewness parameter vector $\lambda$, i.e., $\boldsymbol{Y}|U=u \sim SN_{p}(\boldsymbol{\mu},u^{-1}\boldsymbol{\Sigma},\boldsymbol{\lambda})$. \\
The marginal pdf of $\boldsymbol{Y}_J$ is:
\begin{equation}
f(\boldsymbol{y}_J)=2 \int_0^{\infty} \phi_{p}(\boldsymbol{y}_J|\boldsymbol{\mu},u^{-1}\boldsymbol{\Sigma}) \Phi(u^{1/2} \boldsymbol{\lambda^{T}\Sigma^{-1/2}(\boldsymbol{y}_J-\boldsymbol{\mu})}) dH(u|\boldsymbol{\nu})
\end{equation}

where $\phi_{p}(.|\boldsymbol{\mu},\boldsymbol{\Sigma})$
stands for the pdf of the p-variate normal distribution with mean vector $\boldsymbol{\mu}$ and covariance matrix $\boldsymbol{\Sigma}$, $\Phi(.)$ represents the cdf of the standard univariate normal distribution. Thus, we can say, $\boldsymbol{Y}_J \sim SNI_{p}(\boldsymbol{\mu} , \boldsymbol{\Sigma}, \boldsymbol{\lambda},H)$. \\

	\par The normal/ independent distribution (N/I) family is a class of distributions including the major distributions such as multivariate skew normal, multivariate skew t, multivariate skew slash and multivariate skew contaminated. The last three distributions constitute the class of asymmetric SNI distributions. Different selections of $U$ leads to specific cases of skew independent distributions. The selection of $U\sim\Gamma(\nu/2, \nu/2)$ leads to Student's t distribution, $U\sim Beta(\nu, a)$ to slash distributions and
\begin{eqnarray}
	h(u|\boldsymbol{\nu})&=& \gamma, ~~~~~~~~~~u=\lambda \\
	&=& 1-\gamma, ~~~~~u=1
\end{eqnarray}
where $0<\lambda<1$ and $0\leq \gamma<1$ leads to skew contaminated distributions for $\boldsymbol{Y}_J$. These type of distributions are appropriate for parametric inference where the dataset contains outlying observations or observations which makes the distributional assumptions depart from normality. Thus these distributions can be used as a tool for robust inference as their tails are heavier than the tails of the normal distribution. \\

\textit{\textbf{Lemma 1}} \\
\par Let $\boldsymbol{Y}_{J} \sim SNI_p(\boldsymbol{\mu},\boldsymbol{\Sigma},\boldsymbol{\lambda};H)$ and $\boldsymbol{Y}_J$ is partitioned as $\boldsymbol{Y}_J^{\prime}=(\boldsymbol{Y}_{J1}^{\prime},\boldsymbol{Y}_{J2}^{\prime})^{\prime}$ of dimensions $p_1$ and $p_2$ $(p_1+p_2=p)$, respectively; let $\mu=(\mu_1^{T},\mu_2^{T})^{T}$ and
\begin{equation}
\boldsymbol{\Sigma}= \left(\begin{array}{cc}
\boldsymbol{\Sigma}_{11}&\boldsymbol{\Sigma}_{12}\\ \boldsymbol{\Sigma}_{21}&\boldsymbol{\Sigma}_{22}
\end{array}\right),
\end{equation}
be the corresponding partitions of $\boldsymbol{\mu}$ and $\boldsymbol{\Sigma}$. Then, marginally $\boldsymbol{Y}_{J1} \sim SNI_{p_1} (\boldsymbol{\mu}_1,\boldsymbol{\Sigma}_{11},\boldsymbol{\Sigma}_{11}^{1/2}\overset{\sim}{\boldsymbol{v}};H)$, \\
where $\overset{\sim}{\boldsymbol{v}}=\frac{\boldsymbol{v}_1+\boldsymbol{\Sigma}_{11}^{-1}\boldsymbol{\Sigma}_{12}\boldsymbol{v}_2}{\sqrt{1+\boldsymbol{v}_2^{\prime}\boldsymbol{\Sigma}_{22.1}\boldsymbol{v}_2}}$ \\
with $\boldsymbol{\Sigma}_{22.1}=\boldsymbol{\Sigma}_{22}-\boldsymbol{\Sigma}_{21}\boldsymbol{\Sigma}_{11}^{-1}\boldsymbol{\Sigma}_{12}$ and $\boldsymbol{v}=\boldsymbol{\Sigma}^{-1/2}\boldsymbol{\lambda}=(\boldsymbol{v}_1^{\prime},\boldsymbol{v}_2^{\prime})^{\prime}$.

\vspace{0.8cm}

\textit{\textbf{Lemma 2}}\\
\par Let $\boldsymbol{Y}_J \sim SNI_p(\boldsymbol{\mu},\boldsymbol{\Sigma},\boldsymbol{\lambda};H)$. Then the distribution of $\boldsymbol{Y}_{J2}$, conditionally on $\boldsymbol{Y}_{J1}=\boldsymbol{y}_{J1}$ and $\boldsymbol{U}=\boldsymbol{u}$, has density \\
\begin{equation}
f(\boldsymbol{y}_2|\boldsymbol{y}_1,\boldsymbol{u})= \phi_{p_2}(\boldsymbol{y}_2|\boldsymbol{\mu}_{2.1},\boldsymbol{u}^{-1}\boldsymbol{\Sigma}_{22.1}) \frac{\Phi_{1}(\boldsymbol{u}^{1/2}\boldsymbol{v}^{\prime}(\boldsymbol{y}_J-\boldsymbol{\mu}))}{\Phi_{1}(\boldsymbol{u}^{1/2}\overset{\sim}{\boldsymbol{v}}^{\prime}(\boldsymbol{y}_{J1}-\boldsymbol{\mu}_1))}
\end{equation}
with $\boldsymbol{\mu}_{2.1}=\boldsymbol{\mu}_{2}+\boldsymbol{\Sigma}_{21}\boldsymbol{\Sigma}_{11}^{-1}(\boldsymbol{y}_{J1}-\boldsymbol{\mu}_1)$. Furthermore, we get
\begin{equation}
E[\boldsymbol{Y}_{J2}|\boldsymbol{y}_{J1},\boldsymbol{u}]=\boldsymbol{\mu}_{2.1}+\boldsymbol{u}^{-1/2} \frac{\phi_{1}(\boldsymbol{u}^{1/2}\overset{\sim}{\boldsymbol{v}}^{\prime}(\boldsymbol{y}_{J1}-\boldsymbol{\mu}_1))}{\Phi_{1}(\boldsymbol{u}^{1/2}\overset{\sim}{\boldsymbol{v}}^{\prime}(\boldsymbol{y}_{J1}-\boldsymbol{\mu}_1))} \frac{\boldsymbol{\Sigma}_{22.1}\boldsymbol{v}_2}{\sqrt{1+\boldsymbol{v}_2^{\prime}\boldsymbol{\Sigma}_{22.1}\boldsymbol{v_2}}}
\end{equation}

	\section{Joint model with structural association}
\label{sec:mod2}
\subsection{Model Formulation}

	Let $Y_{ij}$ be the longitudinal observation corresponding to the $i^{th}$ individual $j^{th}$ time point where $i=1,...,m$ and $j=1,..n_i$. Let $T_i$ be the time-to-event for the $i^{th}$ individual. Now, the longitudinal observation for the $i^{th}$ individual is expressed as a linear mixed-effects (LME) model\\

 \begin{equation}
 \boldsymbol{Y}_i= \boldsymbol{X}_i \boldsymbol{\beta}+ \boldsymbol{Z}_{1i} \boldsymbol{b}_i+\boldsymbol{\epsilon}_i
 \end{equation}
 where, for $i=1,2,..,m$ we have
 \begin{equation}
 \left(\begin{array}{c}
 \boldsymbol{b}_i\\\boldsymbol{\epsilon}_i
 \end{array}\right) \overset{iid}{\sim} SNI_{n_i+q} \left( \left(\begin{array}{c} c \boldsymbol{\Delta} \\ \boldsymbol{0} \end{array} \right),     \boldsymbol{\Sigma}=\left(\begin{array}{cc}
 \boldsymbol{D}_b& \boldsymbol{0} \\
\boldsymbol{0} & \boldsymbol{D}_e
 \end{array}\right),  \left(\begin{array}{c}
 \boldsymbol{\lambda}\\ \boldsymbol{0}
 \end{array} \right), \boldsymbol{\nu}       \right)
 \label{jt_ran}
 \end{equation}

 \par Hence, $\boldsymbol{b}_i|\boldsymbol{D}_b, \boldsymbol{\lambda}, \boldsymbol{\nu} \overset{iid}{\sim}SNI_q(\boldsymbol{c}\boldsymbol{\Delta},\boldsymbol{D}_{b},\boldsymbol{\lambda},\boldsymbol{\nu})$ and $\boldsymbol{\epsilon}_i|\sigma^2_e, \boldsymbol{\nu} \overset{ind}{\sim} NI_{n_i}(\boldsymbol{0},\boldsymbol{D}_{e},\boldsymbol{\nu})$
 Here, $\boldsymbol{D}_{b}=\boldsymbol{D(\Omega)}$ is the $q \times q$ dispersion matrix of $\boldsymbol{b}_i$ depending on the unknown reduced parameter vector $\Omega$ and $\boldsymbol{D}_{e}=\sigma^2_e \boldsymbol{I}_{n_i}$, where $\sigma^2_e$ is the variability for the within subjects and $\boldsymbol{\lambda}$ denotes the skewness parameter for the random-effects $\boldsymbol{b}_i$. Again $\boldsymbol{\Delta}=\boldsymbol{D}^{1/2} \boldsymbol{\kappa}$ with $\boldsymbol{\kappa}=\frac{\boldsymbol{\lambda}}{\sqrt{1+\boldsymbol{\lambda}^\prime\boldsymbol{\lambda}}}$ and $c= - \sqrt{2/\pi} E(U_i^{-1/2})$ for the $i^{th}$ subject where $U_i$ signifies the mixing variable having cdf $H(u_i|\boldsymbol{\nu})$. $\boldsymbol{X}_i$ denotes the $n_i \times p$ design matrix corresponding to the fixed effects for the $i^{th}$ subject. Further, $\boldsymbol{\beta}$ and $\boldsymbol{Z}_{1i}$ respectively denote the vector of regression coefficients termed as fixed-effects and the $n_i \times q$ design matrix which corresponds to the $q \times 1$ vector of random-effects $\boldsymbol{b}_i$. Again, $\boldsymbol{\epsilon}_i$ is the $n_i \times 1 $ vector of random errors. Again, it may be noted that $E(\boldsymbol{b}_i|\boldsymbol{D}_b, \boldsymbol{\lambda}, \boldsymbol{\nu})=E(\boldsymbol{\epsilon}_i|\sigma^2_e,\boldsymbol{\nu})=\boldsymbol{0}$ which implies that marginal distribution of the within-subject random-errors are symmetric around zero and the random-effects are asymmetric having mean zero. From equation (\ref{jt_ran}), it is evident that $\boldsymbol{b}_i$ and $\boldsymbol{\epsilon}_i$ are uncorrelated as $Cov(\boldsymbol{b}_i, \boldsymbol{\epsilon}_i| \sigma^2_e, \boldsymbol{D}_b, \boldsymbol{\lambda}, \boldsymbol{\nu})=E(\boldsymbol{b}_i \boldsymbol{\epsilon}_i^\prime|\sigma^2_e, \boldsymbol{D}_b, \boldsymbol{\lambda}, \boldsymbol{\nu})=E[E(\boldsymbol{b}_i \boldsymbol{\epsilon}_i^\prime|\sigma^2_e, \boldsymbol{D}_b, \boldsymbol{\lambda}, u_i)   | \sigma^2_e, \boldsymbol{D}_b, \boldsymbol{\lambda}, \boldsymbol{\nu}]  =\boldsymbol{0}. $
 \\

 Here, $T_i$ denoting the time-to-event observation for the $i^{th}$ individual is taken as the minimum value of the observed value for the $i^{th}$ individual say, $T_i^*$ and the censoring time $C_i$ so as to incorporate the censoring in the time-to-event model. So, $T_i=min(T_i^*,C_i)$ and $\delta_i=0$ if the $i^{th}$ individual is censored and zero otherwise. The observed time-to-event observations finally constitute of $(T_i, \delta_i)$ for the $i^{th}$ subject. The time-to-event is modelled by fully parametric accelerated failure time (AFT) model which is very effective when we are concerned with the acceleration in disease progression based on the time-to-event subdmodel. \\

 \par Again, in this work, we have explored the aspect that the time-to-event process can be directly dependent on the longitudinal process along with the subject-specific random-effects, and hence we have developed a conditional model where the two submodels i.e., the longitudinal and time-to-event are bound together by structural association apart from being linked by the latent variable in form of subject-specific random-effects.
 \vspace{0.5cm}

 Now, let us define the combined data as $\boldsymbol{J}_i=(\boldsymbol{Y}_i^\prime, \log T_i)^\prime$. An $(n_i+1)$-dimensional random vector $\boldsymbol{J}_i$ follows a SNI distribution with location parameter $\boldsymbol{\mu} \in \Re^{n_i+1}$, scale matrix $\boldsymbol{\Sigma}$ (an $(n_i+1) \times (n_i+1$) positive-definite matrix) and skewness parameter $\boldsymbol{\lambda} \in \Re^{n_i+1}$ if its probability density function is given by
\begin{eqnarray}
f(\boldsymbol{J}_i)=2\int_0^\infty \phi_{n_i+1}(\boldsymbol{J}_i|\boldsymbol{\mu}, \boldsymbol{u}^{-1}\boldsymbol{\Sigma}) \Phi(u^{1/2}\boldsymbol{\lambda}_J^\prime \boldsymbol{\Sigma}^{-1/2}(\boldsymbol{J}_i-\boldsymbol{\mu} )) dH(u)
\end{eqnarray}
where $U$ is a positive random variable with cumulative distribution function $H(u|\boldsymbol{\nu})$ and we use the notation $\boldsymbol{J}_i \sim SNI_{n_i+1}(\boldsymbol{\mu} , \boldsymbol{\Sigma}, \boldsymbol{\lambda}_J|H)$.
\\ Given the subject-specific random-effects $\boldsymbol{b}_i$ we assume
\begin{eqnarray*}
\boldsymbol{J}_i=\left(\begin{array}{c}
\boldsymbol{Y}_i\\ \log T_i
\end{array}\right) \sim SNI_{n_i+1} \left(\left(\begin{array}{c}
\boldsymbol{X}_i \boldsymbol{\beta}\\ \beta_{0}
\end{array}\right), \boldsymbol{\Sigma} =\left(\begin{array}{cc}
\boldsymbol{\psi}_i&\boldsymbol{\sigma}\\ \boldsymbol{\sigma}^{\prime}&\sigma_T^2
\end{array}\right),\boldsymbol{\lambda}_J=\left(\begin{array}{c}
\boldsymbol{\lambda}_i \\ 0
\end{array}\right), H \right)
\end{eqnarray*}

where $Cov(\log T_i,\boldsymbol{Y}_i)=\boldsymbol{\sigma}$. Here, $\boldsymbol{\sigma}$ signifies the parametric structural association between the $i^{th}$ longitudinal and time-to-event observation. \\

\par Hence according to Lemma 1, the marginal density of $Y_i$ can be expressed as $SNI_{n_i}(\boldsymbol{X}_i \boldsymbol{\beta},\boldsymbol{\psi}_i, \boldsymbol{\psi}_i^{1/2} \overset{\sim}{\boldsymbol{v}},  H_{\boldsymbol{\nu}})$ where $\overset{\sim}{\boldsymbol{v}}=\frac{\boldsymbol{v}_1+\boldsymbol{\psi}_i^{-1}\boldsymbol{\sigma} \boldsymbol{v}_2}{\sqrt{1+\boldsymbol{v}_2^{\prime}\boldsymbol{\Sigma}_{22.1}\boldsymbol{v}_2}}$, $\Sigma_{22.1}=\sigma_T^2-\boldsymbol{\sigma}^\prime \boldsymbol{\psi}_i^{-1} \boldsymbol{\sigma}$, $\boldsymbol{v}=\boldsymbol{\Sigma}^{-1/2}\boldsymbol{\lambda} =(\boldsymbol{v}_1^\prime,\boldsymbol{v}_2^\prime)^\prime$, $\boldsymbol{\psi}_i=\sigma^2_e\boldsymbol{I}+\boldsymbol{Z}_i \boldsymbol{D}_b \boldsymbol{Z}_i^\prime$,  $\bar{\boldsymbol{\lambda}}_i=\frac{\boldsymbol{\psi}_i^{-1/2} \boldsymbol{Z}_i \boldsymbol{D}_b \boldsymbol{\zeta}}{\sqrt{1+\boldsymbol{\zeta}^\prime \boldsymbol{\Lambda}_i\boldsymbol{\zeta}}}$ where $\boldsymbol{\Lambda}_i=(\boldsymbol{D}_{b}^{-1}+\sigma_e^{-2}\boldsymbol{Z}_i^{\prime}\boldsymbol{I}^{-1}_{n_i} \boldsymbol{Z}_i)^{-1}$ and $\boldsymbol{\zeta}=\boldsymbol{D}_b^{-1/2} \boldsymbol{\lambda}  $. \\

\par Again, the distribution of $\log T$, conditionally on $\boldsymbol{Y}_i=\boldsymbol{y}_i$, $\boldsymbol{b}_i$ and $U=u$, has density
\begin{equation*}
f(\log T_i|\boldsymbol{Y}_{i},\boldsymbol{b}_i,u_i)= \phi(\log T|\mu_{2.1},u^{-1}\Sigma_{22.1}) \frac{\Phi\left(u^{1/2}\boldsymbol{v}^\prime \left(\begin{array}{c}
	\boldsymbol{Y}_i-\boldsymbol{X}_i \beta\\ \log T_i-\beta_0
	\end{array}\right)   \right)}{\Phi(u^{1/2}\overset{\sim}{\boldsymbol{v}}^\prime (\boldsymbol{Y}_{i}-\boldsymbol{X}_{i}\boldsymbol{\beta}))}
\end{equation*}
where $\mu_{2.1}=\beta_0+\boldsymbol{Z}_{2i}\boldsymbol{b}_i+\boldsymbol{\sigma}^\prime \boldsymbol{\psi}_i^{-1} (\boldsymbol{Y}_i-\boldsymbol{X}_i \beta) $ and $\Sigma_{22.1}$ denotes the conditional mean and dispersion matrix of $\log T_i$ given $\boldsymbol{Y}_i$, $\boldsymbol{b}_i$ and $U$. Again, $\boldsymbol{Z}_{2i}$ signifies the $n_i \times q $ design matrix for the $q \times 1$ vector of random-effects for the time-to-event submodel and $\boldsymbol{\sigma}=\sigma^2_{cov} \boldsymbol{1}$. \\

\vspace{0.2cm}
\par In this article, we attempt to develop robust inference of joint modelling of longitudinal and time-to-event data using skewed versions of normal, Student's t, slash and the contaminated normal distribution. We consider a linear mixed-effects model with normal/ independent distributional assumptions for modelling the longitudinal submodel and an accelerated failure time setup for the time-to-event submodel. The processes are linked by structural association as well as the association by the subject-specific random-effects. Bayesian approach is adopted for the model implementation and the analysis was carried out using OpenBUGS. \\

	\subsection{Bayesian approach}
\label{sec:Bayes}
Bayesian framework is adopted using MCMC techniques for the joint modelling of longitudinal and survival data under skewed distributional assumptions.
A key feature of these kinds of models is that they can be fitted with a flexible hierarchical representation which facilitates computation in OpenBugs.
The random variable representing the $(n_i+1)$ variate joint data for the $i^{th}$ individual is $\boldsymbol{J}_i=(\boldsymbol{Y}_i, \log T_i)^\prime$. This observation for joint data is split into two segments to express each submodel namely longitudinal and time-to-event into hierarchical representations.

\begin{eqnarray}
    	\log T_i|\boldsymbol{Y}_i, \boldsymbol{b}_i, U_i=u_i &\overset{ind}{\sim}& N(\mu_{2.1}, \sigma_T^2-\boldsymbol{\sigma}^\prime \boldsymbol{\psi}_i^{-1} \boldsymbol{\sigma}) \label{hier:eq1}\\
	\boldsymbol{Y}_i|\boldsymbol{b}_i,U_i=u_i &\overset{ind}{\sim}& N_{n_i}(\boldsymbol{X}_i \boldsymbol{\beta}+\boldsymbol{Z}_{1i} \boldsymbol{b}_i, u_i^{-1} \sigma^2_e \boldsymbol{I}_{n_i}) \\
	\boldsymbol{b}_i|\boldsymbol{D}_b, u_i, \boldsymbol{\lambda} &\overset{ind}{\sim}& N_{q} (c\boldsymbol{\Delta}, u_i^{-1} \boldsymbol{D}_b) \\
	u_i|\boldsymbol{\nu}    &\overset{iid}{\sim}& h(u_i|\boldsymbol{\nu}) \label{hier:eqL}
\end{eqnarray}

where $\mu_{2.1}=\beta_0+\boldsymbol{Z}_{2i}\boldsymbol{b}_i+\boldsymbol{\sigma}^\prime \boldsymbol{\psi}_i^{-1} (\boldsymbol{Y}_i-\boldsymbol{X}_i \boldsymbol{\beta})$. Hence, the complete data likelihood is given by:
\begin{eqnarray*}
	\mathscr{L}(\boldsymbol{\Theta}|\mathscr{D}) &\propto&  \prod_{i=1}^{m} \phi_{n_i} (\boldsymbol{Y}_i|\boldsymbol{X}_i \boldsymbol{\beta}+\boldsymbol{Z}_{1i} \boldsymbol{b}_i ; u_i^{-1} \sigma^2_e \boldsymbol{I}_{n_i} )  \times  [\phi(\log T|\mu_{2.1},u^{-1}\Sigma_{22.1}) \\
	&\times& \frac{\Phi\left(u^{1/2}\boldsymbol{v}^\prime \left(\begin{array}{c}
			\boldsymbol{Y}_i-\boldsymbol{X}_i \beta\\ \log T_i-\beta_0
		\end{array}\right)   \right)}{\Phi(u^{1/2}\overset{\sim}{\boldsymbol{v}}^\prime (\boldsymbol{Y}_{i}-\boldsymbol{X}_{i}\boldsymbol{\beta}))}]^{\delta_i} \times [\int_{t}^\infty  \phi(\log T|\mu_{2.1},u^{-1}\Sigma_{22.1}) \\
		&\times&  \frac{\Phi\left(u^{1/2}\boldsymbol{v}^\prime \left(\begin{array}{c}
		  		\boldsymbol{Y}_i-\boldsymbol{X}_i \beta\\ \log T_i-\beta_0
		  	\end{array}\right)   \right)}{\Phi(u^{1/2}\overset{\sim}{\boldsymbol{v}}^\prime (\boldsymbol{Y}_{i}-\boldsymbol{X}_{i}\boldsymbol{\beta}))}] \times \phi_{q}(\boldsymbol{b}_i|c\boldsymbol{\Delta},\boldsymbol{D}_b) \times h(u_i|\boldsymbol{\nu})
	\end{eqnarray*}

\par where $\mathscr{D}=\mbox{Data},\Theta=(\boldsymbol{\beta},\beta_0,\sigma_e^{2},\sigma^2_T,\boldsymbol{\sigma}^2,\boldsymbol{D}_b,\boldsymbol{\lambda^\prime},\boldsymbol{\nu^\prime})^\prime$ where $\boldsymbol{Z}_{2i}=\boldsymbol{\nu}^\prime$. \\

\par Prior distributions for the elements of the parameter vector $\Theta$ is considered for completing the Bayesian specification. Proper and conditionally conjugate priors were considered to ensure that the posterior distribution attains standard form. The elements of the parameter vector are considered to be independent and the priors are chosen in a noninformative manner. The parameter vector for the fixed effects for both the submodels are taken as multivariate normal choice for $\boldsymbol{\beta}$ and normal for $\beta_0$. Cauchy prior \citep{gelman2006prior} is considered for the within subject error variability parameter $\sigma^2_e$ and the components of subject-specific dispersion matrix $\boldsymbol{D}_b$. Cauchy priors are specially suitable when considerable information is required on the variability parameters in hierarchical setup. Here, $\boldsymbol{\Delta}$ is chosen as $\Delta \boldsymbol{1}_q$ where $\boldsymbol{1}_q$ signifies a vector with dimension $q$ and each entry 1. A Student's t prior $t(s_1, s_2, s_3)$ suggested by Bayes and Branco \citep{bayes2007bayesian} was taken for $\Delta$ where $s_1$, $s_2$ and $s_3$ respectively denotes the centering parameter, scale and degrees of freedom. This is nothing but an approximate for the Jeffrey's prior for the skewness parameter of the skew-normal distribution. Further, $\boldsymbol{\nu}$ has different prior formulations for different members of the SNI class. The skew-t model has hierarchical prior for $\boldsymbol{\nu}$ i.e., $\nu|\lambda_0 \sim Texp(\lambda_0;(2,\infty))$ and $\lambda_0 \sim U(\lambda_1, \lambda_2)$ (see \citep{cabral2012multivariate}), where $Texp(\lambda, (2, \infty))$ signifies exponential distribution with mean $\frac{1}{\lambda}$ being truncated in the interval (a, b) for the purpose of achieving finite variance. The skew-slash model has $\nu|\lambda_0 \sim exp(\lambda_0)$ and the skew contaminated normal model has $\nu \sim Beta(\nu_0, \nu_1)$ and $\rho \sim Beta(\rho_0, \rho_1)$ in order to achieve conjugacy. \\

The elements of the parameter space $\Theta$ are assumed to be independent among themselves and thus the joint prior distribution for all the unknown parameters can be expressed as $\pi(\Theta)=\pi(\boldsymbol{\beta}) \pi(\beta_0) \pi(\sigma^2_e) \pi(\sigma^2_T) \pi(\boldsymbol{\sigma}^2)\pi(\boldsymbol{D}_b) \pi(\boldsymbol{\lambda})$. Hence, finally the joint posterior distribution likelihood can be expressed as:\\
\begin{eqnarray*}
\mathscr{L}(\mathscr{D}|\boldsymbol{\Theta}) &\propto& \mathscr{L}(\boldsymbol{\Theta}|\mathscr{D}) \pi(\Theta) \\
 &\propto& \prod_{i=1}^{m} \phi_{n_i} (\boldsymbol{Y}_i|\boldsymbol{X}_i \boldsymbol{\beta}+\boldsymbol{Z}_{1i} \boldsymbol{b}_i ; u_i^{-1} \sigma^2_e \boldsymbol{I}_{n_i} )  \times  [\phi(\log T_i|\mu_{2.1},u^{-1}\Sigma_{22.1}) \\
 &\times& \frac{\Phi\left(u^{1/2}\boldsymbol{v}^\prime \left(\begin{array}{c}
 		\boldsymbol{Y}_i-\boldsymbol{X}_i \beta\\ \log T_i-\beta_0
 	\end{array}\right)   \right)}{\Phi(u^{1/2}\overset{\sim}{\boldsymbol{v}}^\prime (\boldsymbol{Y}_{i}-\boldsymbol{X}_{i}\boldsymbol{\beta}))}]^{\delta_i} \times [\int_{t}^\infty  \phi(\log T_i|\mu_{2.1},u^{-1}\Sigma_{22.1}) \\
 &\times&  \frac{\Phi\left(u^{1/2}\boldsymbol{v}^\prime \left(\begin{array}{c}
 		\boldsymbol{Y}_i-\boldsymbol{X}_i \beta\\ \log T_i-\beta_0
 	\end{array}\right)   \right)}{\Phi(u^{1/2}\overset{\sim}{\boldsymbol{v}}^\prime (\boldsymbol{Y}_{i}-\boldsymbol{X}_{i}\boldsymbol{\beta}))}] \times \phi_{q}(\boldsymbol{b}_i|c\boldsymbol{\Delta},\boldsymbol{D}_b) \times h(u_i|\boldsymbol{\nu}) \\
   &&\times \pi(\boldsymbol{\beta}) \times \pi(\beta_0) \times \pi(\sigma^2_e) \times \pi(\sigma^2_T) \times \pi(\boldsymbol{\sigma}^2) \times \pi(\boldsymbol{D}_b) \times \pi(\boldsymbol{\lambda})
\end{eqnarray*}

Again, it is evident that $\mathscr{L}(\mathscr{D}|\boldsymbol{\Theta})$ intractable analytically and hence MCMC methods of sampling such as Metropolis-Hastings or Gibbs sampler is implemented with the help of the hierarchical representation (\ref{hier:eq1})-(\ref{hier:eqL}) to draw samples for obtaining inference.

\section{Simulation Study}
\label{sec:sim}
We conduct an extensive simulation study to compare the performance of the following joint models: skew-normal ($JM_{SN}$), skew-t ($JM_{ST}$), skew-slash ($JM_{SSL}$) and skew-contaminated ($JM_{SN}$) in presence of 5\% and 10\% outliers in the continuous outcomes. We have compared these skewed joint models with the regular non-skewed joint model ($JM$) in order to examine the efficacy of these models in terms of handling influential observations. For each setup of the competing joint models we had generated 100 data sets each having 200 subjects ($m$) with varying number of longitudinal observations ($n_i$). This has been achieved through generation from Uniform(4, 10). A single continuous outcome is used for simulation. Again, 5\% and 10\% outlier observations are incorporated in this covariate to investigate the efficiency of the distributions in terms of robustness.  Covariate values $x_{ij}^{'}$s are generated from N(3, 0.5) distribution. Subject-specific random-effects i.e., $\boldsymbol{b}_i=(b_{i1}, b_{i2})^\prime$ are assumed to be distributed as skew-normal with parameters $c \boldsymbol{\Delta}= 1.5 c \boldsymbol{1}_2$, $\boldsymbol{D}_b= \left(\begin{array}{cc}
\Omega_1&\Omega_2\\
\Omega_2&\Omega_1\\
\end{array}\right)$ where $\Omega_1=1$ and $\Omega_2=0.5$, $\boldsymbol{\lambda}= 1.1~\boldsymbol{1}_2$. Longitudinal observations $y_{ij}^{'s}$ are generated as:\\
\begin{equation}
y_{ij}= \alpha+ \beta x_{ij}+ b_{i1}+t_{ij} b_{i2}+ \epsilon_{ij}
\end{equation}
where $\alpha=$0.9, $\beta=$1 and $\epsilon_{ij}\sim IN(0, \sigma^2_e, 3)$.  \\

Time-to-event observations $\log T_i^s$s are generated from a Frechet family or type II extreme value distribution which is a heavy tailed distribution in the generalized extreme value (GEV) family of distributions\\
\begin{eqnarray*}
	 g(x)=\frac{1}{\sigma} t(x)^{\xi+1} e^{-t(x)},~\mbox{where}~   t(x)&=& (1+\xi (\frac{x-\mu}{\sigma}))^{-\frac{1}{\xi}}~ \mbox{if}~ \xi\neq 0\\
	                                                                &=& e^{-\frac{(x-\mu)}{\sigma}}~ ~~~~~~~~~~~~~~~\mbox{if}~ \xi= 0
\end{eqnarray*}
where $\mu$ is the location parameter and $\sigma$ is the variability/scale parameter. In this set up, we have used
where $\mu=\beta_{0}+\nu_1 b_{i1}+\nu_2 b_{i2}+\sigma_{cov}^{2\prime} \sum_{j=1}^{n_i} (\sigma^2_e+\Omega_1+2 t_{ij} \Omega_2+ t_{ij}^2 \Omega_1)^{-1} (y_{ij}-\alpha -\beta x_{ij}-b_{i1}  -\boldsymbol{t}_{ij}b_{i2})$ as the location parameter and $\sigma_{22.1}=(\sigma^2_{T}- 	\boldsymbol{\sigma}^{2\prime}_{cov}	(\sigma^2_e+\Omega_1+2 t_{ij} \Omega_2+ t_{ij}^2 \Omega_1)^{-1} 	\boldsymbol{\sigma}^2_{cov})$ as the variability/ scale parameter. $\xi$ is taken as 0.8 to ensure the heavy tailed nature which is efficient in incorporating extreme values. Two levels of outliers i.e., 5\% and 10\% has been considered for the sake of comparison and the results are summarized in Tables \ref{GEV1} and \ref{GEV2}. Also, in another setting we have generated $\log T_{i}^{'}s$ from normal with the same mean and variability parameters and compared with respect to 5\% and 10\% levels of outliers. The findings are displayed in Tables \ref{norm1} and \ref{norm2}.\\

\par Here, $\beta_0=$1, $\nu_1=$0.8, $\nu_2=$0.9, $\sigma^2_{cov}=$0.3, $\sigma^2_e=$0.5 and $\sigma^2_T=$0.99. Censoring times $C_i^{'}$s are drawn from an exponential distribution with mean 0.5 and $T_i^*= \min(\log T_i, C_i)$. The censoring percentage has been observed to be around 40\%. \\

\par The longitudinal submodel had been fitted by fitting the competing models differing in random error and random-effects distribution keeping in mind both the skewness in covariate distributions and the possible presence of outliers in the data set. These models are:\\

1. Skew-normal (SN): Independent multivariate skew-normal density for random-effects and random-errors \\
2. Skew-t (ST): Independent multivariate skew-t density for random-effects and random-errors \\
3. Skew-Slash (SSL): Independent multivariate skew-slash density for random-effects and random-errors \\
4. Skew-Contaminated (SCN): Independent multivariate skew-contaminated normal density for random-effects and random-errors \\

For the priors in Bayesian model framework, we have adopted conjugate normal priors for fixed-effects parameters for both the submodels i.e., $\boldsymbol{\beta}_1$ and $\beta_0$ with high degree of precision. For the variability components truncated cauchy priors were considered, where $\sigma^2_e \sim Cauchy(0, 25)$ , $\Omega_1 \sim Cauchy(0, 25)$, $\Omega_2 \sim Cauchy(0, 25)$. The prior for $\boldsymbol{\Delta}$ is taken as $t(0, 25, 2)$. We take $\nu \sim TExp(\lambda_{0})$ with $\lambda_{0} \sim U(a,b)$ for the skew-t model, $\nu \sim Exp(\lambda_{0})$ with $\lambda_{0} \sim U(c, d)$ for the skew-slash model and $\nu \sim Beta(1,1)$ and $\rho \sim Beta(2,2)$ for the skew-contaminated normal model. Here the values of a, b, c, d are chosen that the expected value of $\nu$ lies in the interval $[2, 50]$.  Bayesian trace plots and the autocorrelation plots has been examined from the summary to ensure the desired number of burn-in iterations to be 5000 and we have examined the MCMC convergence and the mixing of the chains. We also examined three parallel MCMC chains with overdispersed initial values and the inference has been based on 5000 iteration values for each chain. The chains exhibit rapid convergence, all achieving convergence within 5000 iterations with all the parameters displaying $\hat{R}$ (the between-chain and within-chain standard deviation ratio) to be around 1.01. The estimation results reported in Tables \ref{GEV1}, \ref{GEV2}, \ref{norm1} and \ref{norm2}, display the bias (the deviation of the estimated value from the true parameter value), standard deviation (SD) of the estimated posterior means) standard error (SE), i.e., the square root of the average of the posterior variance, and the coverage probabilities (CP) of 95\% equal-tailed credible intervals (CI). It may be observed that though JM model gives satisfactory result in terms of bias for the parameters involved in longitudinal sub-model ($\alpha$ and $\tau_z$), the CP values are very low. This low CP may be attributed to the higher SE values. As the percentage of outliers increases, for most of the parameters, in general, biases increase along with decrease in CP values (Tables \ref{GEV2} and \ref{norm2}).
\\


\section{Data Analysis}
\label{sec:data}
\subsection{AIDS Data}

\begin{figure}[H]
    \centering
    \subfloat{{\includegraphics[width=5.5cm, height=8cm]{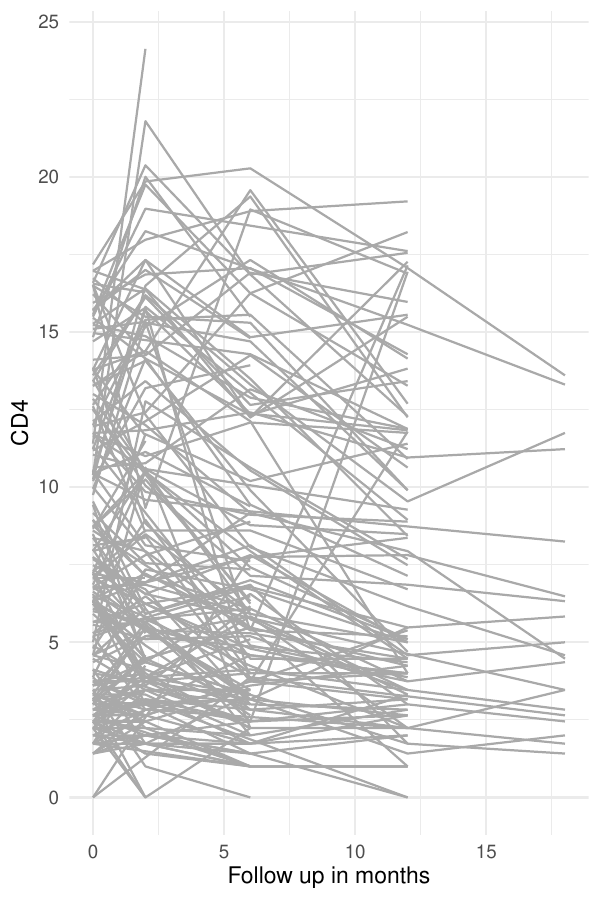} }}
    \qquad
    \subfloat{{\includegraphics[width=5.5cm, height=8cm]{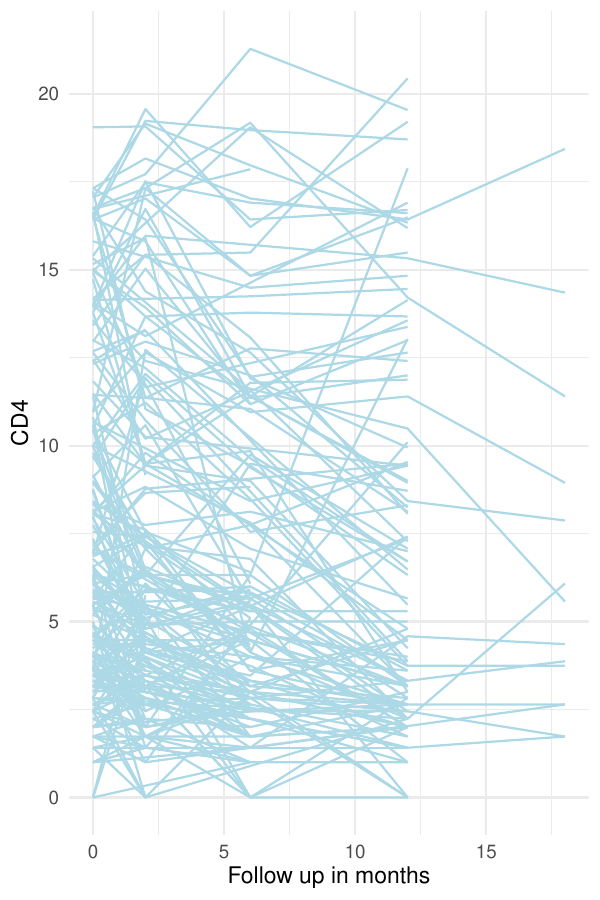} }}
    \caption{Profile plot of 467 patients under (a) Didanosine (b) Zalcitabine}
    \label{fig:profile}
\end{figure}

\begin{figure}[H]
    \centering
    \subfloat{{\includegraphics[width=5cm, height=8cm]{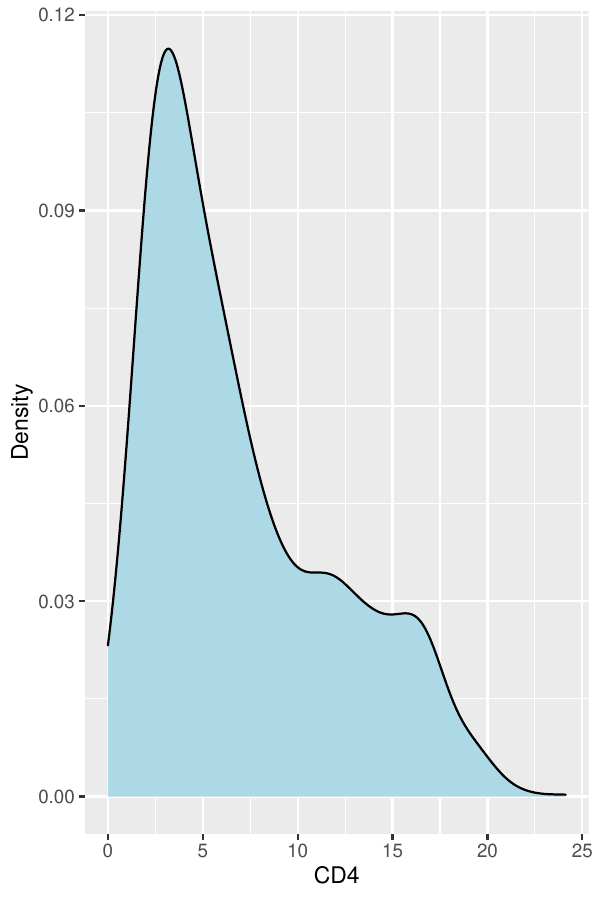} }}
    \qquad
    \subfloat{{\includegraphics[width=5cm, height=8cm]{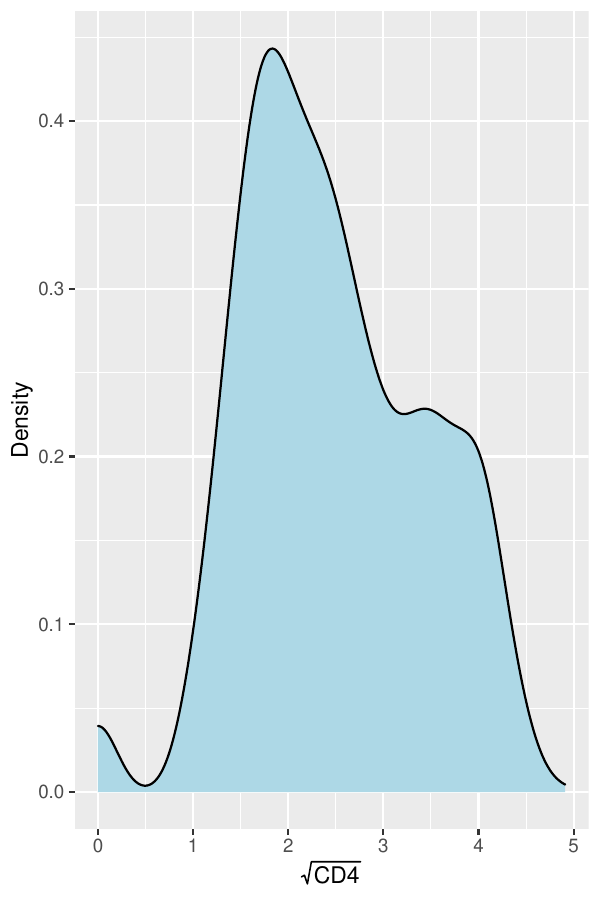} }}
    \caption{(a) Density plot of CD4 cell count. (b) Density plot of $\sqrt{CD4}$ cell count}
    \label{fig:den}
\end{figure}

The data consists of 467 HIV-infected patients having enrolled in ddI/ ddC study under Terry Beirn Community Programs for Clinical Research on AIDS. The patients either had symptoms of AIDS or had CD4 cell counts below 300 cells per cubic millimetre. 230 patients were randomly assigned with didanosine (ddI) at a dosage of 500mg/day and 237 patients received zalcitabine (ddC) at a dosage of 2.25mg/day. The goal of this randomized open label clinical trial was to compare the effectiveness of the two alternative antiretroviral drugs i.e., ddI and ddC in HIV affected patients who had experienced adverse side effects under Zidovudine (ZDV) or had progression of the disease in spite of taking ZDV. A detailed description and discussion about this dataset is available in Goldman et al. \citep{goldman1996response}. Zidovudine was primarily used as an antiretroviral drug and was effective in delaying the occurrence of opportunistic infections in patients and enhanced the condition of survival. However, ZDV was expensive and patients experienced conditions like nausea, vomiting, diarrhoea and problems with the nervous system and pancreas. Here, in this study, CD4 cell counts were recorded at study entry and for the subsequent 2, 6, 12 and 18 months thereafter. Figure \ref{fig:profile} represents the profile plot of 467 patients under each drug type. The plots indicate highly unbalanced nature of the CD4 cell counts over follow up time. The density plot in Figure \ref{fig:den} (a) depicts the left-skewed nature of the CD4 cell counts. \\

\par
\begin{figure}[H]
    \centering
    \includegraphics[width=6cm, height=6cm]{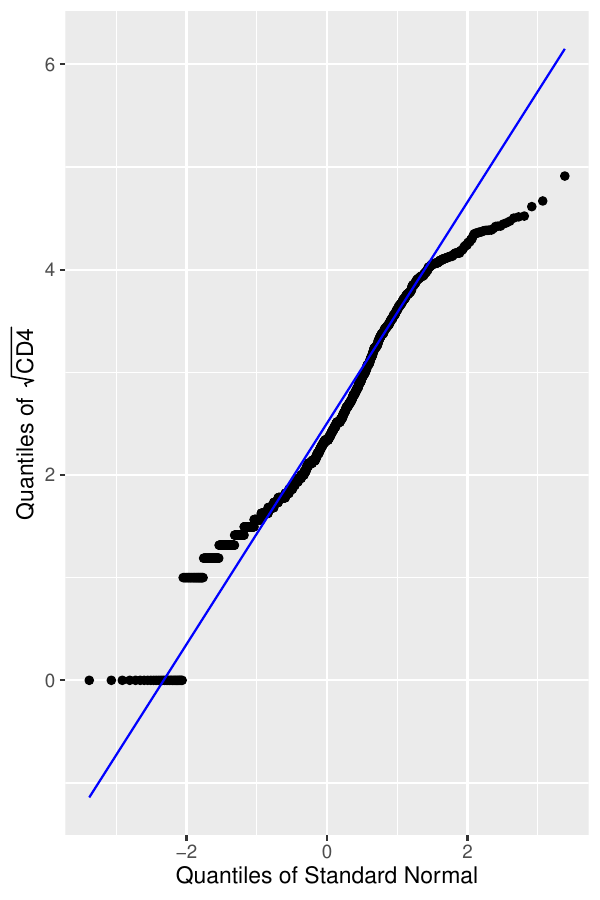}
    \caption{QQ plot of $\sqrt{CD4}$} 
    \label{QQplot}
\end{figure}

 The skewness in CD4 cell data indicates the fact that normality assumption is not very appealing for this dataset. The usual method adopted to address this issue is square root transformation such that the data can be fitted under the normality framework. However, we can clearly see from Figure \ref{fig:den} (b) that this does not remove the skewed nature of the CD4 cell count completely. This fact is supported by Figure \ref{QQplot}. Hence, normality assumption in this case lacks robustness. We analyze this data set using the proposed joint modelling framework using robust distributions.

\par The CD4 cell counts clearly acts as a biomarker for the progression of AIDS and hence indicates the better or worsened immunity condition in patients. The CD4 trajectories for each patient in the study is expressed as the mixed-effects model with information regarding Observation time, Gender, intolerance to Zidovudine therapy, type of drug administered, previous opportunistic infection incorporated as covariates in the model. The covariates are coded in the form of binary explanatory variables i.e., Gender (1=male, -1=female), AZT-Status (-1=intolerance, 1=failed), Drug (0=ddC, 1=ddI), Previous Opportunistic Infection, i.e., PrevOI (1=AIDS diagnosis at baseline, 0=no AIDS diagnosis at baseline) as the influencing factors for the CD4 cell count in the peripheral blood. Hence, we 

\begin{eqnarray*}
	Y_{ij}&=& \beta_{11}+\beta_{12}\mbox{Obstime}_{ij}+\beta_{13} \mbox{Drug}_{i}+\beta_{14}\mbox{Gender}_{i}+\beta_{15}\mbox{PrevOI}_{i} \\
	&&+\beta_{16} \mbox{AZT-Status}_{i}+b_{i1}+b_{i2}\mbox{Obstime}_{ij}+\epsilon_{1ij} \\
\end{eqnarray*}

where $Y_{ij}$ denotes the square root of the CD4 cell count for the $i^{th}$ patient at the $j^{th}$ time point with $\epsilon_{1ij}\sim IN(0,\sigma^2_e, H_{\nu})$. Here, the vector of regression coefficients for the longitudinal trajectory is $\boldsymbol{\beta}_1=(\beta_{11}, \beta_{12}, \beta_{13}, \beta_{14}, \beta_{15}, \beta_{16})^T$. Again, according to our model, the time-to-event outcome variable i.e., death due to acquired immunodeficiency syndrome can been expressed as:

\begin{eqnarray*}
\log T_i&=& \beta_{0}+\nu_1 b_{i1}+\nu_2 b_{i2}\\
&&+\boldsymbol{\sigma}_{cov}^{2\prime} \sum_{j=1}^{n_i} (\sigma^2_e+\Omega_1+2  \Omega_2 \mbox{Obstime}_{ij}+ \Omega_1  \mbox{Obstime}_{ij}^2)^{-1}\times \\
&& (Y_{ij}-\beta_{11}-\beta_{12}\mbox{Obstime}_{ij}-\beta_{13}~ \mbox{Drug}_{i}-\beta_{14}\mbox{Gender}_{i}-\beta_{15}\mbox{PrevOI}_{i} \\
&&-\beta_{16} \mbox{AZT-Status}_{i}-b_{i1}-b_{i2}\mbox{Obstime}_{ij})+\epsilon_{2ij}
\end{eqnarray*}
where $\log T_i$ denotes the time-to-event observation indicating the time upto which $i^{th}$ patient has survived (if death occurs) or time at which the observation was censored (if the patient was alive at the end of the study), $\beta_0$ being the intercept of the time-to-event trajectory, $\nu_1$ and $\nu_2$ the coefficients of the intercept and slope components of the random-effects and $\epsilon_{2ij}\sim N(0,\tau_t^{-1})$. $\sigma^2_{cov}$ captures the structural covariance that is present between the CD4 cell count observations of a patient and the time to death or progression in disease severity. \\

\par The joint modelling framework adopted here highlights the structural association between the two submodels i.e., we assume that the CD4 cell count is directly responsible for the increase or decrease in disease progression of the patient. Hence, the covariates contributing to the CD4 cell count observations are also contributing to the time to death or disease progression of the patient and hence it is sufficient to include the covariates in the longitudinal submodel only. We have cross-checked the fact by including the covariates (those capable of influencing the time-to-event directly) in both submodels and have obtained similar results. \\

Bayesian inference is used to find the parameter estimates using OpenBugs. For the prior selection, same prior distributions are considered. Three parallel MCMC chains are setup with 5000 as burn-in samples and 5000 as iteration samples with over-dispersed initial values. The trace plots and autocorrelation plots have been investigated and they reveal no potential sign of non-convergence for any of the parameters. Moreover, $\hat{R}$ for all the model parameters are seen to be within 1.1. \\

These distributions as described earlier provide support against anomaly in inference due to outlier observations. The time-to-event submodel has been developed using the fully parametric accelerated failure time model where a censored observation means that the patient was found to be alive at the end of the study period. Results displayed in Table \ref{tab:aidsskew} show comparable results for all the parameters except the longitudinal and time-to-event precision i.e., $\tau_z$ and $\tau_t$ where the estimates are not much comparable. $\Omega_1$ also displays a great discrepancy in the estimates.

\section{Discussion} \label{sec:dissSNI}
Here, in this work we have deviated from the traditional joint modelling techniques in two ways i.e., in terms of improvement in the association between the two processes when subject-specific random-effects are given, and in terms of dealing with influential observations. Here, structural association between the two processes ensures that the two submodels remain attached even if the subject-specific random-effects is given. This is achieved by means of the conditional distributional assumptions where we have $T| \boldsymbol{Y}, \boldsymbol{b}$ where $T$ is the time-to-event observation, $\boldsymbol{Y}$ the longitudinal observation and $\boldsymbol{b}$ the subject-specific random-effect. Hence, apart from being dependent on the shared subject-specific random-effects, the time-to-event process also depends on the longitudinal process directly. This makes our joint model more general than the existing joint models where processes become independent when subject-specific random-effects are given. Again, the adoption of the robust distribution in the parametric setup takes care of the influential observations. This fact has been examined by a detailed simulation study by incorporating cases of 5\% and 10\% outliers. The study reveals that the joint models using the class of skew-normal/independent distributions are highly efficient in dealing with a heavy influence of extreme observations in the data as opposed to the joint model with linear mixed-effects. As Bayesian approach is used for inference, a sensitivity analysis with varying choices of priors can be carried out to examine the efficacy of this robust class of joint models.

\section*{Acknowledgements}

\bibliographystyle{biom}
\bibliography{Ref_skew_mod}

\begin{thebibliography}{}

\bibitem[\protect\citeauthoryear{Andrews and Mallows}{Andrews and
  Mallows}{1974}]{andrews1974scale}
Andrews, D.~F. and Mallows, C.~L. (1974).
\newblock Scale mixtures of normal distributions.
\newblock {\em Journal of the Royal Statistical Society: Series B
  (Methodological)} {\bf 36,} 99--102.

\bibitem[\protect\citeauthoryear{Bandyopadhyay, Castro, Lachos, and
  Pinheiro}{Bandyopadhyay et~al.}{2015}]{bandyopadhyay2015robust}
Bandyopadhyay, D., Castro, L.~M., Lachos, V.~H., and Pinheiro, H.~P. (2015).
\newblock Robust joint non-linear mixed-effects models and diagnostics for
  censored hiv viral loads with cd4 measurement error.
\newblock {\em Journal of Agricultural, Biological, and Environmental
  Statistics} {\bf 20,} 121--139.

\bibitem[\protect\citeauthoryear{Bayes and Branco}{Bayes and
  Branco}{2007}]{bayes2007bayesian}
Bayes, C.~L. and Branco, M.~D. (2007).
\newblock Bayesian inference for the skewness parameter of the scalar
  skew-normal distribution.
\newblock {\em Brazilian Journal of Probability and Statistics} pages 141--163.

\bibitem[\protect\citeauthoryear{Brilleman, Crowther, Moreno-Betancur, Buros,
  Dunyak, Al-Huniti, and et~al.}{Brilleman et~al.}{2019}]{brillman81}
Brilleman, S., Crowther, M., Moreno-Betancur, M., Buros, N.~J., Dunyak, J.,
  Al-Huniti, N., and et~al. (2019).
\newblock Joint longitudinal and time-to-event models for multilevel
  hierarchical data.
\newblock {\em Stat Methods Med Res} {\bf 28(12),} 3502–15.

\bibitem[\protect\citeauthoryear{Brown and Ibrahim}{Brown and
  Ibrahim}{2003}]{brown2003bayesian}
Brown, E.~R. and Ibrahim, J.~G. (2003).
\newblock A bayesian semiparametric joint hierarchical model for longitudinal
  and survival data.
\newblock {\em Biometrics} {\bf 59,} 221--228.

\bibitem[\protect\citeauthoryear{Cabral, Lachos, and Prates}{Cabral
  et~al.}{2012}]{cabral2012multivariate}
Cabral, C. R.~B., Lachos, V.~H., and Prates, M.~O. (2012).
\newblock Multivariate mixture modeling using skew-normal independent
  distributions.
\newblock {\em Computational Statistics \& Data Analysis} {\bf 56,} 126--142.

\bibitem[\protect\citeauthoryear{Chi and Ibrahim}{Chi and
  Ibrahim}{2006}]{chi2006joint}
Chi, Y.-Y. and Ibrahim, J.~G. (2006).
\newblock Joint models for multivariate longitudinal and multivariate survival
  data.
\newblock {\em Biometrics} {\bf 62,} 432--445.

\bibitem[\protect\citeauthoryear{Dutta, Molenberghs, and Chakraborty}{Dutta
  et~al.}{2021}]{dutta2021joint}
Dutta, S., Molenberghs, G., and Chakraborty, A. (2021).
\newblock Joint modelling of longitudinal response and time-to-event data using
  conditional distributions: a bayesian perspective.
\newblock {\em Journal of Applied Statistics} pages 1--18.

\bibitem[\protect\citeauthoryear{Faucett and Thomas}{Faucett and
  Thomas}{1996}]{faucett1996simultaneously}
Faucett, C.~L. and Thomas, D.~C. (1996).
\newblock Simultaneously modelling censored survival data and repeatedly
  measured covariates: a gibbs sampling approach.
\newblock {\em Statistics in medicine} {\bf 15,} 1663--1685.

\bibitem[\protect\citeauthoryear{Fitzmaurice, Davidian, Verbeke, and
  Molenberghs}{Fitzmaurice et~al.}{2008}]{fitzmaurice2008longitudinal}
Fitzmaurice, G., Davidian, M., Verbeke, G., and Molenberghs, G. (2008).
\newblock {\em Longitudinal data analysis}.
\newblock CRC press.

\bibitem[\protect\citeauthoryear{Gelman et~al\mbox{.}}{Gelman
  et~al.}{2006}]{gelman2006prior}
Gelman, A. et~al. (2006).
\newblock Prior distributions for variance parameters in hierarchical models
  (comment on article by browne and draper).
\newblock {\em Bayesian analysis} {\bf 1,} 515--534.

\bibitem[\protect\citeauthoryear{Goldman, Carlin, Crane, Launer, Korvick,
  Deyton, and Abrams}{Goldman et~al.}{1996}]{goldman1996response}
Goldman, A.~I., Carlin, B.~P., Crane, L.~R., Launer, C., Korvick, J.~A.,
  Deyton, L., and Abrams, D.~I. (1996).
\newblock Response of cd4 lymphocytes and clinical consequences of treatment
  using ddi or ddc in patients with advanced hiv infection.
\newblock {\em JAIDS Journal of Acquired Immune Deficiency Syndromes} {\bf 11,}
  161--169.

\bibitem[\protect\citeauthoryear{Huang, Li, and Elashoff}{Huang
  et~al.}{2010}]{huang2010joint}
Huang, X., Li, G., and Elashoff, R.~M. (2010).
\newblock A joint model of longitudinal and competing risks survival data with
  heterogeneous random effects and outlying longitudinal measurements.
\newblock {\em Statistics and its interface} {\bf 3,} 185.

\bibitem[\protect\citeauthoryear{K and S.}{K and S.}{2017}]{li2017}
K, L. and S., L. (2017).
\newblock Dynamic predictions in bayesian functional joint models for
  longitudinal and time-to-event data: an application to alzheimer’s disease.
\newblock {\em Stat Methods Med Res} {\bf 28,} 1–16.

\bibitem[\protect\citeauthoryear{Lange, Little, and Taylor}{Lange
  et~al.}{1989}]{lange1989robust}
Lange, K.~L., Little, R.~J., and Taylor, J.~M. (1989).
\newblock Robust statistical modeling using the t distribution.
\newblock {\em Journal of the American Statistical Association} {\bf 84,}
  881--896.

\bibitem[\protect\citeauthoryear{Li and Luo}{Li and Luo}{2019}]{liluo2019}
Li, K. and Luo, S. (2019).
\newblock Bayesian functional joint models for multivariate longitudinal and
  time-to-event data.
\newblock {\em Comput Stat Data Anal} {\bf 129,} 14–29.

\bibitem[\protect\citeauthoryear{Li, Elashoff, and Li}{Li
  et~al.}{2009}]{li2009robust}
Li, N., Elashoff, R.~M., and Li, G. (2009).
\newblock Robust joint modeling of longitudinal measurements and competing
  risks failure time data.
\newblock {\em Biometrical Journal: Journal of Mathematical Methods in
  Biosciences} {\bf 51,} 19--30.

\bibitem[\protect\citeauthoryear{Lin, McCulloch, and Mayne}{Lin
  et~al.}{2002}]{lin2002maximum}
Lin, H., McCulloch, C.~E., and Mayne, S.~T. (2002).
\newblock Maximum likelihood estimation in the joint analysis of time-to-event
  and multiple longitudinal variables.
\newblock {\em Statistics in Medicine} {\bf 21,} 2369--2382.

\bibitem[\protect\citeauthoryear{Little}{Little}{1993}]{little1993pattern}
Little, R.~J. (1993).
\newblock Pattern-mixture models for multivariate incomplete data.
\newblock {\em Journal of the American Statistical Association} {\bf 88,}
  125--134.

\bibitem[\protect\citeauthoryear{Little and Rubin}{Little and
  Rubin}{1989}]{little1989analysis}
Little, R.~J. and Rubin, D.~B. (1989).
\newblock The analysis of social science data with missing values.
\newblock {\em Sociological Methods \& Research} {\bf 18,} 292--326.

\bibitem[\protect\citeauthoryear{Little and Rubin}{Little and
  Rubin}{2000}]{little2000causal}
Little, R.~J. and Rubin, D.~B. (2000).
\newblock Causal effects in clinical and epidemiological studies via potential
  outcomes: concepts and analytical approaches.
\newblock {\em Annual review of public health} {\bf 21,} 121--145.

\bibitem[\protect\citeauthoryear{Mchunu, Mwambi, Rizopoulos, Reddy, and
  Yende-Zuma}{Mchunu et~al.}{2022}]{mchunu2022using}
Mchunu, N.~N., Mwambi, H.~G., Rizopoulos, D., Reddy, T., and Yende-Zuma, N.
  (2022).
\newblock Using joint models to study the association between cd4 count and the
  risk of death in tb/hiv data.
\newblock {\em BMC Medical Research Methodology} {\bf 22,} 1--9.

\bibitem[\protect\citeauthoryear{Molenberghs, Thijs, Kenward, and
  Verbeke}{Molenberghs et~al.}{2003}]{molenberghs2003sensitivity}
Molenberghs, G., Thijs, H., Kenward, M.~G., and Verbeke, G. (2003).
\newblock Sensitivity analysis of continuous incomplete longitudinal outcomes.
\newblock {\em Statistica Neerlandica} {\bf 57,} 112--135.

\bibitem[\protect\citeauthoryear{Pawitan and Self}{Pawitan and
  Self}{1993}]{pawitan1993modeling}
Pawitan, Y. and Self, S. (1993).
\newblock Modeling disease marker processes in aids.
\newblock {\em Journal of the American Statistical Association} {\bf 88,}
  719--726.

\bibitem[\protect\citeauthoryear{Pinheiro, Liu, and Wu}{Pinheiro
  et~al.}{2001}]{pinheiro2001efficient}
Pinheiro, J.~C., Liu, C., and Wu, Y.~N. (2001).
\newblock Efficient algorithms for robust estimation in linear mixed-effects
  models using the multivariate t distribution.
\newblock {\em Journal of Computational and Graphical Statistics} {\bf 10,}
  249--276.

\bibitem[\protect\citeauthoryear{Rosa, Padovani, and Gianola}{Rosa
  et~al.}{2003}]{rosa2003robust}
Rosa, G., Padovani, C.~R., and Gianola, D. (2003).
\newblock Robust linear mixed models with normal/independent distributions and
  bayesian mcmc implementation.
\newblock {\em Biometrical Journal: Journal of Mathematical Methods in
  Biosciences} {\bf 45,} 573--590.

\bibitem[\protect\citeauthoryear{Rustand, van Niekerk, Krainski, Rue, and
  Proust-Lima}{Rustand et~al.}{2023}]{rustand2023fast}
Rustand, D., van Niekerk, J., Krainski, E.~T., Rue, H., and Proust-Lima, C.
  (2023).
\newblock Fast and flexible inference for joint models of multivariate
  longitudinal and survival data using integrated nested laplace
  approximations.
\newblock {\em Biostatistics} page kxad019.

\bibitem[\protect\citeauthoryear{Sousa}{Sousa}{2011}]{sousa}
Sousa, I. (2011).
\newblock A review on joint modelling of longitudinal measurements and
  time-to-event.
\newblock {\em Revstat Stat J} {\bf 9,} 57--81.

\bibitem[\protect\citeauthoryear{Tseng, Hsieh, and Wang}{Tseng
  et~al.}{2005}]{tseng2005joint}
Tseng, Y.-K., Hsieh, F., and Wang, J.-L. (2005).
\newblock Joint modelling of accelerated failure time and longitudinal data.
\newblock {\em Biometrika} {\bf 92,} 587--603.

\bibitem[\protect\citeauthoryear{Verbeke and Lesaffre}{Verbeke and
  Lesaffre}{1996}]{verbeke1996linear}
Verbeke, G. and Lesaffre, E. (1996).
\newblock A linear mixed-effects model with heterogeneity in the random-effects
  population.
\newblock {\em Journal of the American Statistical Association} {\bf 91,}
  217--221.

\bibitem[\protect\citeauthoryear{Wang, Luo, and Li}{Wang et~al.}{2017}]{wang17}
Wang, J., Luo, S., and Li, L. (2017).
\newblock Dynamic prediction for multiple repeated measures and event time
  data: an application to parkinson’s.
\newblock {\em Ann Appl Stat.} {\bf 11(3),} 1787–809.

\bibitem[\protect\citeauthoryear{Wang and Taylor}{Wang and
  Taylor}{2001}]{wang2001jointly}
Wang, Y. and Taylor, J. M.~G. (2001).
\newblock Jointly modeling longitudinal and event time data with application to
  acquired immunodeficiency syndrome.
\newblock {\em Journal of the American Statistical Association} {\bf 96,}
  895--905.

\bibitem[\protect\citeauthoryear{Wu}{Wu}{2009}]{wu2009mixed}
Wu, L. (2009).
\newblock {\em Mixed effects models for complex data}.
\newblock Chapman and Hall/CRC.

\bibitem[\protect\citeauthoryear{Wulfsohn and Tsiatis}{Wulfsohn and
  Tsiatis}{1997}]{wulfsohn1997joint}
Wulfsohn, M.~S. and Tsiatis, A.~A. (1997).
\newblock A joint model for survival and longitudinal data measured with error.
\newblock {\em Biometrics} pages 330--339.

\bibitem[\protect\citeauthoryear{Xu and Zeger}{Xu and
  Zeger}{2001}]{xu2001joint}
Xu, J. and Zeger, S.~L. (2001).
\newblock Joint analysis of longitudinal data comprising repeated measures and
  times to events.
\newblock {\em Journal of the Royal Statistical Society: Series C (Applied
  Statistics)} {\bf 50,} 375--387.

\bibitem[\protect\citeauthoryear{Zhang and Davidian}{Zhang and
  Davidian}{2001}]{zhang2001linear}
Zhang, D. and Davidian, M. (2001).
\newblock Linear mixed models with flexible distributions of random effects for
  longitudinal data.
\newblock {\em Biometrics} {\bf 57,} 795--802.

\bibitem[\protect\citeauthoryear{Zou, Zeng, Xiao, and Luo}{Zou
  et~al.}{2023}]{zou2023bayesian}
Zou, H., Zeng, D., Xiao, L., and Luo, S. (2023).
\newblock Bayesian inference and dynamic prediction for multivariate
  longitudinal and survival data.
\newblock {\em The Annals of Applied Statistics} {\bf 17,} 2574--2595.

\end{thebibliography}

\appendix

\begin{table} \tiny
	\setlength{\tabcolsep}{2pt}
	\begin{tabular}{ccccccccccccccccccccc}
		\hline
		\multicolumn{1}{c}{\small{\quad}}&\multicolumn{1}{c}{\small{True}}&\multicolumn{3}{c}{$JM_{SN}$}&\quad&\multicolumn{3}{c}{$JM_{ST}$}&\quad&\multicolumn{3}{c}{$JM_{SSL}$}&\quad&\multicolumn{3}{c}{$JM_{SCN}$}&\quad&\multicolumn{3}{c}{$JM$}\\\cline{3-5} \cline{7-9} \cline{11-13} \cline{15-17} \cline{19-21}
		\quad&\quad & BIAS &SD& SE  &\quad& BIAS & SD& SE &\quad& BIAS & SD& SE &\quad& BIAS & SD& SE &\quad& BIAS & SD& SE \\
		\hline
		\vspace{0.3cm}
		For longitudinal \\
		$\alpha$& 0.9000& -0.2061 & 0.0477& 0.0887& \qquad& -0.2014& 0.0014&0.0883& \qquad&-0.2106  & 0.0365& 0.0870& \qquad& -0.2029 & 0.0407& 0.0878&\qquad&0.7359&0.0190&0.2487\\
		\vspace{0.3cm}
		\qquad& \qquad&  (0.9600)& \qquad& \qquad& \qquad&  (0.9600)& \qquad& \qquad& \qquad&  (0.9600) & & & \qquad&  (0.9700)& & &\qquad&(\textbf{0.4200})&\qquad&\qquad\\
		$\beta$& 1.0000& 0.2804 & 0.1829& 0.0384& \qquad& 0.2045& 0.0145&0.0111& \qquad&0.2321  & 0.0232& 0.0384& \qquad& 0.2317 & 0.0249& 0.0321&\qquad&0.3464&0.0054&0.0750\\
		\vspace{0.3cm}
		\qquad& \qquad&  (0.9200)& \qquad& \qquad& \qquad&  (0.8000)& \qquad& \qquad& \qquad&  (0.8600) & & & \qquad&  (0.8000)& & &\qquad&(\textbf{0.5200})&\qquad&\qquad\\
		
		$\tau_z$& 2.0000& -0.1457 & 0.1004& 0.0984& \qquad& -0.1005& 0.0295&0.0329& \qquad&0.1293  & 0.0983& 0.1759& \qquad& 0.5623 & 0.3169& 0.3729&\qquad&\textbf{-1.1027}&0.0005&0.0463\\
		\qquad& \qquad&  (1.0000)& \qquad& \qquad& \qquad&  (1.0000)& \qquad& \qquad& \qquad&  (1.0000) & & & \qquad&  (1.0000)& & &\qquad&(\textbf{0.4000})&\qquad&\qquad\\
		\vspace{0.3cm}
		For both processes \\
		
		$\Omega_1$& 1.0000& 0.8621& 0.3252& 0.3892&  \qquad& \textbf{1.0235}& 0.1395& 0.1279&  \qquad&\textbf{ 1.0285}&0.4106 & 0.5671&  \qquad& \textbf{1.4263} & 0.2911&0.3682 & &-&-&-\\
		\vspace{0.3cm}
		\qquad& \qquad&  (\textbf{0.4800})& \qquad& \qquad& \qquad&  (0.7900)& \qquad& \qquad& \qquad&  (0.7500)& & & \qquad&  (0.5200)& & &&-\\
		
		$\Omega_2$& 0.5000& -0.0001& 0.0008& 0.0018&  \quad& 0.0002& 0.0005& 0.0005&  \qquad& 0.0001& 0.0008& 0.0998&  \qquad& 0.0003 & 0.0005& 0.0005& \qquad&-&-&-\\
		\vspace{0.3cm}
		\quad& \quad&  (1.0000)& \qquad& \qquad& \qquad&  (1.0000)& \qquad& \qquad& \qquad&  (1.0000)& & & \qquad&  (1.0000)& & &\qquad&-\\
		$\lambda_1$& 1.1000& -0.0921& 0.0045 &0.0717 &  \qquad& \textbf{-1.0674}&0.0007 & 0.0146&  \qquad& \textbf{-0.8813}&0.0601 & 0.1229&  \qquad& -0.0905& 0.0043&0.0057& \qquad&-&-&-\\
		
		\vspace{0.3cm}
		\quad& \quad&  (1.0000)& \qquad& \qquad& \qquad&  (0.5200)& \qquad& \qquad& \qquad&  (0.5000)& & & \qquad&  (1.0000) & & &&-\\
		
		$\lambda_2$& 1.1000& -0.0895& 0.0025& 0.0057&  \qquad& -0.0930& 0.0001&0.0005& \qquad& -0.0914&0.0028& 0.0572& \qquad& -0.0916 &0.0045& 0.0057& \qquad&-&-&-\\
		\vspace{0.3cm}
		\qquad& \qquad&  (1.0000)&\qquad& \qquad& \qquad&  (1.0000)&  \qquad& \qquad& \qquad&  (1.0000)&  & & \qquad&  (1.0000) & & &\qquad&-\\

		$\sigma^2_{cov}$& 0.3000& 0.0001& 0.0011& 0.0173&  \qquad& 0.0008& 0.0003& 0.0737& \qquad& 0.0003& 0.0013& 0.0173&  \qquad& 0.0003 & 0.0013 & 0.0173 & \qquad&0.0068&0.0015&0.1732\\
		\vspace{0.3cm}
		\qquad& \qquad&  (1.0000)&\qquad& \qquad& \qquad& (1.0000)&   \qquad& \qquad& \qquad&  (1.0000)&  & & \qquad&  (1.0000) & & &&(1.0000)\\

		$\sigma^2_{b}$& 0.8000& -&-& -&  \qquad& -& -& -& \qquad& -& -& -&  \qquad& - & - & - & \qquad&0.1736&0.0020&0.0010\\
		\vspace{0.3cm}
		\qquad& \qquad&  -&\qquad& \qquad& \qquad& -&   \qquad& \qquad& \qquad&  -&  & & \qquad&  - & & &&(1.0000)\\
		
		$\rho$& 0.3000& -&-& -&  \qquad& -& -& -& \qquad& -& -& -&  \qquad& - & - & - & \qquad&0.0382&0.0049&0.1142\\
		\vspace{0.3cm}
		\qquad& \qquad&  -&\qquad& \qquad& \qquad& -&   \qquad& \qquad& \qquad&  -&  & & \qquad&  - & & &&(1.0000)\\
		\vspace{0.3cm}
		For time-to-event \\

		$\beta_0$& 1.0000& -0.0001 & 0.0010& 0.0099& \qquad& -0.0013& 0.0009&0.0009& \qquad&-0.0003  & 0.0007& 0.0799& \qquad& -0.0004 & 0.0008& 0.0101&\qquad&\textbf{-0.9175}&2.0394&316.77\\
		\vspace{0.3cm}
		\qquad& \qquad&  (1.0000)& \qquad& \qquad& \qquad&  (1.0000)& \qquad& \qquad& \qquad&  (1.0000) & & & \qquad&  (1.0000)& & &\qquad&(1.0000)&\qquad&\qquad\\
		
		$\nu_1$& 0.8000& 0.2003 & 0.0008& 0.0100& \qquad& 0.1998& 0.0001&0.0998& \qquad&0.1998  & 0.0006& 0.0998& \qquad& 0.1998 & 0.0008& 0.0999&\qquad&\textbf{9.9228}&0.0627&9.9630\\
		\vspace{0.3cm}
		\qquad& \qquad&  (0.8200)& \qquad& \qquad& \qquad&  (0.6500)& \qquad& \qquad& \qquad&  (0.8900) & & & \qquad&  (0.8500)& & &\qquad&(\textbf{0.3400})&\qquad&\qquad\\
		
		$\nu_2$& 0.9000& 0.0999 & 0.0007& 0.0998& \qquad& 0.0999& 0.0001&0.0995& \qquad&0.0999  & 0.0006& 0.0099& \qquad& 0.1003 & 0.0009& 0.0019&\qquad&\textbf{9.9086}&0.0948&9.9928\\
		\vspace{0.3cm}
		\qquad& \qquad&  (1.0000)& \qquad& \qquad& \qquad&  (1.0000)& \qquad& \qquad& \qquad&  (1.0000) & & & \qquad&  (1.0000)& & &\qquad&(\textbf{0.3200})&\qquad&\qquad\\

		$\tau_t$& 1.0100& -0.0099 & 0.0009& 0.0100& \qquad& -0.0110& 0.0004&0.0991& \qquad&-0.0099  & 0.0008& 0.0099& \qquad& -0.0097 & 0.0009& 0.0100&\qquad&-0.0108&0.0787&9.6665\\
		\vspace{0.3cm}
		\qquad& \qquad&  (1.0000)& \qquad& \qquad& \qquad&  (1.0000)& \qquad& \qquad& \qquad&  (1.0000) & & & \qquad&  (1.0000)& & &\qquad&(1.0000)&\qquad&\qquad\\
		\hline
	\end{tabular}
	\caption{Simulation results from models $JM_{SN}$, $JM_{ST}$, $JM_{SSL}$ and $JM_{SCN}$ and $JM$ when time-to-event data generated from generalized extreme value distribution and 5\% outliers in the continuous outcome. Large bias and poor CP are highlighted in bold. CP is indicated within parenthesis under Bias}
	\vspace{0.5cm}

	\label{GEV1}
\end{table}

\begin{table} \tiny
	\vspace{0.5cm}
	\setlength{\tabcolsep}{2pt}
	\begin{tabular}{ccccccccccccccccccccc}
		\hline
		\multicolumn{1}{c}{\small{\quad}}&\multicolumn{1}{c}{\small{True}}&\multicolumn{3}{c}{$JM_{SN}$}&\quad&\multicolumn{3}{c}{$JM_{ST}$}&\quad&\multicolumn{3}{c}{$JM_{SSL}$}&\quad&\multicolumn{3}{c}{$JM_{SCN}$}&\quad&\multicolumn{3}{c}{$JM$}\\\cline{3-5} \cline{7-9} \cline{11-13} \cline{15-17} \cline{19-21}

		\quad&\quad & BIAS &SD& SE  &\quad& BIAS & SD& SE &\quad& BIAS & SD& SE &\quad& BIAS & SD& SE &\quad& BIAS & SD& SE \\
		
		\hline
		\vspace{0.3cm}
		For longitudinal \\
		
		$\alpha$& 0.9000& -0.2213 & 0.0297& 0.0873& \qquad& -0.2175& 0.0873&0.0866& \qquad&-0.2232  & 0.0364& 0.0872& \qquad& -0.2331 & 0.0316& 0.0786&\qquad&0.7240&0.2331& 0.3322\\
		\vspace{0.3cm}
		\qquad& \qquad&  (0.9600)& \qquad& \qquad& \qquad&  (0.9600)& \qquad& \qquad& \qquad&  (0.8000) & & & \qquad&  (0.8900)& & &\qquad&(\textbf{0.4200})&\qquad&\qquad\\
		$\beta$& 1.0000& 0.2231 & 0.0528& 0.0308& \qquad& 0.2490& 0.0314&0.0222& \qquad&0.2307  & 0.0308& 0.0306& \qquad& 0.2472 & 0.0361& 0.0297&\qquad&0.3513&0.0047&0.0012\\
		\vspace{0.3cm}
		\qquad& \qquad&  (0.5800)& \qquad& \qquad& \qquad&  (0.6000)& \qquad& \qquad& \qquad&  (0.6500) & & & \qquad&  (0.6500)& & &\qquad&(\textbf{0.3200})&\qquad&\qquad\\
		
		$\tau_z$& 2.0000& -0.1577 & 0.0934& 0.0984& \qquad& -0.0800& 0.1136&0.1232& \qquad&0.1293  & 0.0983& 0.1759& \qquad& 0.7827 & 0.2233& 0.2893&\qquad&\textbf{-1.1025}&0.0004&0.0001\\
		\qquad& \qquad&  (0.8100)& \qquad& \qquad& \qquad&  (1.0000)& \qquad& \qquad& \qquad&  (1.0000) & & & \qquad&  (1.0000)& & &\qquad&(\textbf{0.3900})&\qquad&\qquad\\
		\vspace{0.3cm}
		For both processes \\
		
		$\Omega_1$& 1.0000& 0.7188& 0.3389& 0.4291&  \qquad& 0.4310& 0.3631& 0.3391&  \qquad&\textbf{0.9217}& 0.4412& 0.4660&  \qquad& \textbf{1.2008} & 0.1928& 0.2637& &-&-&-\\
		\vspace{0.3cm}
		\qquad& \qquad&  (\textbf{0.4800})& \qquad& \qquad& \qquad&  (0.6200)& \qquad& \qquad& \qquad&  (0.5100)& & & \qquad&  (\textbf{0.4800})& & &&-\\
		
		$\Omega_2$& 0.5000& 0.0001& 0.0007& 0.0099&  \qquad& -0.0014& 0.0989& 0.0982&  \qquad& 0.0001& 0.0008& 0.0997&  \qquad& 0.0001 & 0.0005& 0.0005& \qquad&-&-&-\\
		\vspace{0.3cm}
		\quad& \quad&  (1.0000)& \qquad& \qquad& \qquad&  (1.0000)& \qquad& \qquad& \qquad&  (1.0000)& & & \qquad&  (1.0000)& & &\qquad&-\\
		$\lambda_1$& 1.1000& -0.0905& 0.0044 &0.0718 &  \qquad& \textbf{-1.0629}& 0.0212& 0.0210&  \qquad& -0.8472& 0.0813& 0.2281&  \qquad& -0.0914& 0.0032&0.0572& \qquad&-&-&-\\
		
		\vspace{0.3cm}
		\quad& \quad&  (1.0000)& \qquad& \qquad& \qquad&  (\textbf{0.3800})& \qquad& \qquad& \qquad&  (\textbf{0.3200})& & & \qquad&  (1.0000) & & &&-\\
		
		$\lambda_2$& 1.1000& -0.0891& 0.0048& 0.0572&  \qquad& -0.0106& 0.5711&0.0483& \qquad& -0.0892&0.0060& 0.0570& \qquad& -0.0896 &0.0052& 0.0057& \qquad&-&-&-\\
		\vspace{0.3cm}
		\qquad& \qquad&  (1.0000)&\qquad& \qquad& \qquad&  (1.0000)&  \qquad& \qquad& \qquad&  (1.0000)&  & & \qquad&  (1.0000) & & &\qquad&-\\

		$\sigma^2_{cov}$& 0.3000& 0.0008& 0.0094& 0.0173&  \qquad& 0.0017& 0.1737& 0.1269& \qquad& -0.0004& 0.0012& 0.0173&  \qquad& \quad-0.0001 & 0.0017 & 0.0173 & \qquad&0.0004&0.0008&0.1272\\
		\vspace{0.3cm}
		\qquad& \qquad&  (1.0000)&\qquad& \qquad& \qquad& (1.0000)&   \qquad& \qquad& \qquad&  (1.0000)&  & & \qquad&  (1.0000) & & &&(1.0000)\\

		$\sigma^2_{b}$& 0.8000& -&-& -&  \qquad& -& -& -& \qquad& -& -& -&  \qquad& - & - & - & \qquad&0.1734&0.0018&0.0011\\
		\vspace{0.3cm}
		\qquad& \qquad&  -&\qquad& \qquad& \qquad& -&   \qquad& \qquad& \qquad&  -&  & & \qquad&  - & & &&(1.0000)\\
		
		$\rho$& 0.3000& -&-& -&  \qquad& -& -& -& \qquad& -& -& -&  \qquad& - & - & - & \qquad&0.0406&0.0042&0.0044\\
		\vspace{0.3cm}
		\qquad& \qquad&  -&\qquad& \qquad& \qquad& -&   \qquad& \qquad& \qquad&  -&  & & \qquad&  - & & &&(1.0000)\\
		\vspace{0.3cm}
		For time-to-event \\

		$\beta_0$& 1.0000& -0.0001 & 0.0008& 0.0100& \qquad& 0.0100& 0.0998&0.0891& \qquad&-0.0003  & 0.0007& 0.0099& \qquad& -0.0004 & 0.4259& 0.3999&\qquad&0.0067&2.0525&315.46\\
		\vspace{0.3cm}
		\qquad& \qquad&  (1.0000)& \qquad& \qquad& \qquad&  (1.0000)& \qquad& \qquad& \qquad&  (1.0000) & & & \qquad&  (1.0000)& & &\qquad&(1.0000)&\qquad&\qquad\\
		
		$\nu_1$& 0.8000& 0.2008 &0.0010& 0.0999& \qquad& 0.2000& 0.0995&0.0983& \qquad&0.1997  & 0.0007& 0.0999& \qquad& 0.1994 & 0.0005& 0.0100&\qquad&\textbf{9.9200}&0.0816&9.0983\\
		\vspace{0.3cm}
		\qquad& \qquad&  (0.8900)& \qquad& \qquad& \qquad&  (0.6500)& \qquad& \qquad& \qquad&  (0.8400) & & & \qquad&  (0.7500)& & &\qquad&(\textbf{0.3400})&\qquad&\qquad\\
		
		$\nu_2$& 0.9000& 0.0999 & 0.0009& 0.0099& \qquad& 0.0997& 0.0994&0.0996& \qquad&0.1000  & 0.0007& 0.0999& \qquad& 0.1000 & 0.0008& 0.0019&\qquad&\textbf{9.9072}&0.0853&9.0763\\
		\vspace{0.3cm}
		\qquad& \qquad&  (1.0000)& \qquad& \qquad& \qquad&  (1.0000)& \qquad& \qquad& \qquad&  (1.0000) & & & \qquad&  (1.0000)& & &\qquad&(\textbf{0.3000})&\qquad&\qquad\\

		$\tau_t$& 1.0100& -0.0100 & 0.0067& 0.0100& \qquad& -0.0105& 0.0996&0.0995& \qquad&-0.0103  & 0.0013& 0.0098& \qquad& -0.0101 & 0.0005& 0.0100&\qquad&-0.0227&0.0403&9.8770\\
		\vspace{0.3cm}
		\qquad& \qquad&  (1.0000)& \qquad& \qquad& \qquad&  (1.0000)& \qquad& \qquad& \qquad&  (1.0000) & & & \qquad&  (1.0000)& & &\qquad&(1.0000)&\qquad&\qquad\\
		\hline
	\end{tabular}
		\caption{Simulation results from models $JM_{SN}$, $JM_{ST}$, $JM_{SSL}$ and $JM_{SCN}$ and $JM$ when time-to-event data generated from generalized extreme value distribution and 10\% outliers in the continuous outcome. Large bias and poor CP are highlighted in bold. CP is indicated within parenthesis under Bias}
	\label{tab:sim4SNI}
	\label{GEV2}
\end{table}

\begin{table} \tiny
	\vspace{0.5cm}
	\setlength{\tabcolsep}{2pt}
	\begin{tabular}{ccccccccccccccccccccc}
		\hline
		\multicolumn{1}{c}{\small{\quad}}&\multicolumn{1}{c}{\small{True}}&\multicolumn{3}{c}{$JM_{SN}$}&\quad&\multicolumn{3}{c}{$JM_{ST}$}&\quad&\multicolumn{3}{c}{$JM_{SSL}$}&\quad&\multicolumn{3}{c}{$JM_{SCN}$}&\quad&\multicolumn{3}{c}{$JM$}\\\cline{3-5} \cline{7-9} \cline{11-13} \cline{15-17} \cline{19-21}

		\quad&\quad & BIAS &SD& SE  &\quad& BIAS & SD& SE &\quad& BIAS & SD& SE &\quad& BIAS & SD& SE &\quad& BIAS & SD& SE \\
		
		\hline
		\vspace{0.3cm}
		For longitudinal \\
		
		$\alpha$& 0.9000& 0.1832 & 0.0827& 0.0539& \quad& -0.0058 & 0.1138& 0.0320& \quad& 0.0659  & 0.0816& 0.0578& \quad& 0.0606 & 0.0839& 0.0426&\quad&-0.0028&0.1427&0.1217\\
		\vspace{0.3cm}
		\quad& \quad&  (1.0000)& \quad& \quad& \quad&  (1.0000)& \quad& \quad& \quad&  (0.9600) & & & \quad&  (1.0000)& & &\quad&(0.9400)&\quad&\quad\\
		$\beta$& 1.0000& 0.0657&0.0264& 0.0245&  \quad& 0.0669& 0.0366& 0.0295&  \quad& 0.0556& 0.0263& 0.0247&  \quad& 0.0525 & 0.0291& 0.0384& \quad&0.0036&0.0177&0.0163\\
		\vspace{0.3cm}
		\quad& \quad&  (1.0000)& \quad& \quad& \quad&  (1.0000)& \quad& \quad& \quad&  (1.0000)& & & \quad&  (1.0000)& & &&(0.9800)&\quad\\
		
		$\tau_z$& 2.0000& -0.1364& 0.1012& 0.1058& \quad& 0.3006& 0.3652& 0.1775&   \quad& 0.1145& 0.1777& 0.1401&  \quad&  1.7061 &2.1015 &2.1310 &\quad&-0.0707&0.0426&0.0396\\
		\quad& \quad&  (1.0000)& \quad& \quad& \quad&  (1.0000)& \quad& \quad& \quad&  (0.9600)& & & \quad&  (0.9800) & & &&(0.5000)\\
		\vspace{0.3cm}
		For both processes \\
		
		$\Omega_1$& 1.0000& \textbf{0.5421}& 0.3428& 0.3452&  \quad& -0.0486& 0.9814& 0.2959&  \quad&\textbf{ 0.7144}& 0.6618& 0.4555&  \quad& \textbf{1.6890} & 2.8770& 1.4560& &-&-&-\\
		\vspace{0.3cm}
		\quad& \quad&  (0.6900)& \quad& \quad& \quad&  (1.0000)& \quad& \quad& \quad&  (0.6500)& & & \quad&  (0.5700)& & &&-\\
		
		$\Omega_2$& 0.5000& 0.0004& 0.1002& 0.0015&  \quad& -0.0003& 0.0999& 0.0019&  \quad& 0.0001& 0.0997& 0.0016&  \quad& 0.0002 & 0.1002& 0.0018& \quad&-&-&-\\
		\vspace{0.3cm}
		\quad& \quad&  (1.0000)& \quad& \quad& \quad&  (1.0000)& \quad& \quad& \quad&  (1.0000)& & & \quad&  (1.0000)& & &\quad&-\\
		$\lambda_1$& 1.1000& -0.0894& 0.5728& 0.0083&  \quad& \textbf{-0.5371}& 0.6132& 0.1604&  \quad& \textbf{-0.8743}& 0.2226& 0.0559&  \quad& -0.0948 & 0.5708& 0.0096& \quad&-&-&-\\
		
		\vspace{0.3cm}
		\quad& \quad&  (1.0000)& \quad& \quad& \quad&  (1.0000)& \quad& \quad& \quad&  (\textbf{0.3400})& & & \quad&  (1.0000) & & &&-\\
		
		$\lambda_2$& 1.1000& -0.0900& 0.5710& 0.0102&  \quad& -0.0907& 0.5707& 0.0095& \quad& -0.0911& 0.5711& 0.0096& \quad& -0.0934 & 0.5719& 0.0095& \quad&-&-&-\\
		\vspace{0.3cm}
		\quad& \quad&  (1.0000)&\quad& \quad& \quad&  (1.0000)&  \quad& \quad& \quad&  (1.0000)&  & & \quad&  (1.0000) & & &\quad&-\\

		$\sigma^2_{cov}$& 0.3000& -0.0003& 0.1730& 0.0025&  \quad& -0.0003& 0.1729& 0.0031& \quad& 0.0001& 0.1731& 0.0029&  \quad& 0.0006 & 0.1733 & 0.0031 & \quad&0.0007&0.1731&0.0026\\
		\vspace{0.3cm}
		\quad& \quad&  (1.0000)&\quad& \quad& \quad& (1.0000)&   \quad& \quad& \quad&  (1.0000)&  & & \quad&  (1.0000) & & &&(1.0000)\\

		$\sigma^2_{b}$& 0.8000& -&-& -&  \quad& -& -& -& \quad& -& -& -&  \quad& - & - & - & \quad&0.0481&0.0978&0.0906\\
		\vspace{0.3cm}
		\quad& \quad&  -&\quad& \quad& \quad& -&   \quad& \quad& \quad&  -&  & & \quad&  - & & &&(1.0000)\\
		
		$\rho$& 0.3000& -&-& -&  \quad& -& -& -& \quad& -& -& -&  \quad& - & - & - & \quad&-0.0190&0.1165&0.1171\\
		\vspace{0.3cm}
		\quad& \quad&  -&\quad& \quad& \quad& -&   \quad& \quad& \quad&  -&  & & \quad&  - & & &&(1.0000)\\
		\vspace{0.3cm}
		For time-to-event \\

		$\beta_0$& 1.0000& -0.0001&0.0099& 0.0014&  \quad& 0.0001& 0.0998& 0.0014& \quad& -0.0002& 0.0999&0.0015&  \quad& -0.0006 & 0.0999& 0.0019& \quad&\textbf{3.560}& 4.9888&3.1603\\
		\vspace{0.3cm}
		\quad& \quad&  (1.0000)&\quad& \quad& \quad& (1.0000)&   \quad& \quad& \quad&  (1.0000)&  & & \quad&  (1.0000) & & &&(0.5100)\\
		
		$\nu_1$& 0.8000& 0.1998& 0.0996&0.0018 &  \quad& 0.2002& 0.0999& 0.0015&  \quad& 0.2002& 0.1000& 0.0019&  \quad& 0.2002 & 0.1003 & 0.0019 & \quad&\textbf{10.020}&0.1375&9.9116\\
		\vspace{0.3cm}
		\quad& \quad&  (0.8700)&\quad& \quad& \quad& (0.7700)&   \quad& \quad& \quad&  (0.9700)&  & & \quad&  (0.8300) & & &\quad&(\textbf{ 0.3400})\\
		
		$\nu_2$& 0.9000& 0.0998& 0.1723& 0.0018& \quad& 0.1000& 0.1731& 0.0018&  \quad& 0.1004& 0.1021& 0.0016&  \quad& 0.0996 & 0.1003 & 0.0021 & \quad&\textbf{9.9172}&0.1231&9.9677\\
		\vspace{0.3cm}
		\quad& \quad&  (1.0000)&\quad& \quad& \quad& (1.0000)&   \quad& \quad& \quad&  (1.0000)&  & & \quad&  (1.0000)  & & &\quad&(\textbf{0.3400})\\

		$\tau_t$& 1.0100& -0.0104& 0.0999& 0.0019& \quad&-0.0100& 0.0997& 0.0018&   \quad& -0.0097& 0.1001& 0.0013&  \quad& -0.0096 & 0.1003 & 0.0022 & \quad& -0.1333&0.3073&9.7188\\
		\quad& \quad&  (1.0000)&\quad& \quad& \quad& (1.0000)&   \quad& \quad& \quad&  (1.0000)&  & & \quad&  (1.0000)  & & &\quad&(1.0000)\\
		\hline
	\end{tabular}
		\caption{Simulation results from models $JM_{SN}$, $JM_{ST}$, $JM_{SSL}$ and $JM_{SCN}$ and $JM$ when time-to-event data generated from normal distribution and 5\% outliers in the continuous outcome. Large bias and poor CP are highlighted in bold. CP is indicated within parenthesis under Bias}
	\label{norm1}
\end{table}

\begin{table} \tiny 
	\vspace{0.5cm}
	\setlength{\tabcolsep}{2pt}
	\begin{tabular}{ccccccccccccccccccccc}
		\hline
		\multicolumn{1}{c}{\small{\quad}}&\multicolumn{1}{c}{\small{True}}&\multicolumn{3}{c}{$JM_{SN}$}&\quad&\multicolumn{3}{c}{$JM_{ST}$}&\quad&\multicolumn{3}{c}{$JM_{SSL}$}&\quad&\multicolumn{3}{c}{$JM_{SCN}$}&\quad&\multicolumn{3}{c}{$JM$}\\\cline{3-5} \cline{7-9} \cline{11-13} \cline{15-17} \cline{19-21}

		\quad&\quad & BIAS &SD& SE  &\quad& BIAS & SD& SE &\quad& BIAS & SD& SE &\quad& BIAS & SD& SE &\quad& BIAS & SD& SE \\
		
		\hline
		\vspace{0.3cm}
		For longitudinal \\
		
		$\alpha$& 0.9000& -0.0143 & 0.0853&0.0394& \quad& -0.0577 & 0.0985& 0.0262& \quad& -0.0236  & 0.0832& 0.0357& \quad& -0.0156 & 0.0865& 0.0310&\quad&0.0248&0.1221&0.1921\\
		\vspace{0.3cm}
		\quad& \quad&  (1.0000)& \quad& \quad& \quad&  (1.0000)& \quad& \quad& \quad&  (1.0000) & & & \quad&  (1.0000)& & &\quad&(0.9000)&\quad&\quad\\
		$\beta$& 1.0000& 0.1593&0.0396& 0.0384&  \quad& 0.1545& 0.0564& 0.0338&  \quad& 0.0556& 0.0263& 0.0247&  \quad& 0.0531 & 0.0298& 0.0391& \quad&-0.0120&0.0170&0.0339\\
		\vspace{0.3cm}
		\quad& \quad&  (0.6100)& \quad& \quad& \quad&  (0.6300)& \quad& \quad& \quad&  (0.7900)& & & \quad&  (0.7500)& & &&(0.8900)&\quad\\
		
		$\tau_z$& 2.0000& -0.1418& 0.1007& 0.1298& \quad& 0.3146& 0.3588& 0.1685&   \quad& 0.1292& 0.1777& 0.1329&  \quad&  \textbf{2.3388} &2.9900 &2.8849 &\quad&-0.1033&0.0399&0.0390\\
		\quad& \quad&  (0.6700)& \quad& \quad& \quad&  (0.9600)& \quad& \quad& \quad&  (0.9600)& & & \quad&  (0.9100) & & &&(\textbf{0.3000})\\
		\vspace{0.3cm}
		For both processes \\
		
		$\Omega_1$& 1.0000& \textbf{0.5413}& 0.3224& 0.2693&  \quad& -0.0299& 0.7416& 0.2109&  \quad&\textbf{ 2.3957}& 3.6592& 2.003&  \quad& \textbf{1.6890} & 2.8770& 1.4560& &-&-&-\\
		\vspace{0.3cm}
		\quad& \quad&  (0.6500)& \quad& \quad& \quad&  (1.0000)& \quad& \quad& \quad&  (\textbf{0.4700})& & & \quad&  (0.5100)& & &&-\\
		
		$\Omega_2$& 0.5000& 0.0000& 0.0999& 0.0019&  \quad& 0.0005& 0.1001& 0.0015&  \quad& 0.0001& 0.1000& 0.0021&  \quad& -0.0002 & 0.0998& 0.0019& \quad&-&-&-\\
		\vspace{0.3cm}
		\quad& \quad&  (1.0000)& \quad& \quad& \quad&  (1.0000)& \quad& \quad& \quad&  (1.0000)& & & \quad&  (1.0000)& & &\quad&-\\
		$\lambda_1$& 1.1000& -0.0887& 0.5721& 0.0096&  \quad& \textbf{-0.5318}& 0.6058& 0.1524&  \quad& \textbf{-0.8694}& 0.2203& 0.0651&  \quad& -0.0885 & 0.5709& 0.0096& \quad&-&-&-\\
		
		\vspace{0.3cm}
		\quad& \quad&  (1.0000)& \quad& \quad& \quad&  (0.9900)& \quad& \quad& \quad&  (\textbf{0.4400})& & & \quad&  (\textbf{0.3400}) & & &&-\\
		
		$\lambda_2$& 1.1000& -0.0911& 0.5729& 0.0112&  \quad& -0.0901& 0.5704& 0.0106& \quad& -0.0909& 0.5706& 0.0099& \quad& -0.0881 & 0.5717& 0.0113& \quad&-&-&-\\
		\vspace{0.3cm}
		\quad& \quad&  (1.0000)&\quad& \quad& \quad&  (1.0000)&  \quad& \quad& \quad&  (1.0000)&  & & \quad&  (1.0000) & & &\quad&-\\

		$\sigma^2_{cov}$& 0.3000& 0.0003& 0.1733& 0.0033&  \quad& 0.0003& 0.1732& 0.0029& \quad& 0.0009& 0.1730& 0.0030&  \quad& 0.0004 & 0.1735 & 0.0033 & \quad&0.0003&0.1735&0.0028\\
		\vspace{0.3cm}
		\quad& \quad&  (1.0000)&\quad& \quad& \quad& (1.0000)&   \quad& \quad& \quad&  (1.0000)&  & & \quad&  (1.0000) & & &&(1.0000)\\

		$\sigma^2_{b}$& 0.8000& -&-& -&  \quad& -& -& -& \quad& -& -& -&  \quad& - & - & - & \quad&0.0138&0.0931&0.1240\\
		\vspace{0.3cm}
		\quad& \quad&  -&\quad& \quad& \quad& -&   \quad& \quad& \quad&  -&  & & \quad&  - & & &&(0.9000)\\
		
		$\rho$& 0.3000& -&-& -&  \quad& -& -& -& \quad& -& -& -&  \quad& - & - & - & \quad&-0.0005&0.1171&0.0722\\
		\vspace{0.3cm}
		\quad& \quad&  -&\quad& \quad& \quad& -&   \quad& \quad& \quad&  -&  & & \quad&  - & & &&(1.0000)\\
		\vspace{0.3cm}
		For time-to-event \\

		$\beta_0$& 1.0000& -0.00001&0.0999& 0.0016&  \quad& 0.0000& 0.1003& 0.0016& \quad& -0.0002& 0.0999&0.0019&  \quad& -0.0001 & 0.0999& 0.0017& \quad&\textbf{3.8800}&316.71& 4.4485\\
		\vspace{0.3cm}
		\quad& \quad&  (1.0000)&\quad& \quad& \quad& (1.0000)&   \quad& \quad& \quad&  (1.0000)&  & & \quad&  (1.0000) & & &&(0.5000)\\
		
		$\nu_1$& 0.8000& 0.1998& 0.0996&0.0017 &  \quad& 0.2000& 0.0997& 0.0018&  \quad& 0.2003& 0.1000& 0.0019&  \quad& 0.2000 & 0.1005 & 0.0016 & \quad&\textbf{9.9263}&10.038&0.1951\\
		\vspace{0.3cm}
		\quad& \quad&  (0.8300)&\quad& \quad& \quad& (0.7700)&   \quad& \quad& \quad&  (0.8300)&  & & \quad&  (0.8300) & & &\quad&(\textbf{ 0.3400})\\
		
		$\nu_2$& 0.9000& 0.0100& 0.1724& 0.0018& \quad& 0.0996& 0.0997& 0.0018&  \quad& 0.1005& 0.1023& 0.0018&  \quad& 0.1009 & 0.1008 & 0.0035 & \quad&\textbf{9.9114}&10.012&0.1839\\
		\vspace{0.3cm}
		\quad& \quad&  (1.0000)&\quad& \quad& \quad& (1.0000)&   \quad& \quad& \quad&  (1.0000)&  & & \quad&  (1.0000)  & & &\quad&(\textbf{0.3000})\\

		$\tau_t$& 1.0100& -0.0102& 0.1000& 0.0014& \quad&-0.0100& 0.0998& 0.0021&   \quad& -0.0099& 0.1002& 0.0014&  \quad& -0.0099 & 0.1006 & 0.0015 & \quad& -0.1364&9.7898&0.3091\\
		\quad& \quad&  (1.0000)&\quad& \quad& \quad& (1.0000)&   \quad& \quad& \quad&  (1.0000)&  & & \quad&  (1.0000)  & & &\quad&(0.9000)\\
		\hline
	\end{tabular}
\caption{Simulation results from models $JM_{SN}$, $JM_{ST}$, $JM_{SSL}$ and $JM_{SCN}$ and $JM$ when time-to-event data generated from normal distribution and 10\% outliers in the continuous outcome. Large bias and poor CP are highlighted in bold. CP is indicated within parenthesis under Bias.}
\label{norm2}
\end{table}

\begin{table} \tiny 
	\vspace{0.5cm}
	\setlength{\tabcolsep}{2pt}
	\begin{tabular}{ccccccccccccccccccccc}
		\hline
		\multicolumn{1}{c}{\small{\quad}}&\quad&\multicolumn{3}{c}{$JM_{SN}$}&\quad&\multicolumn{3}{c}{$JM_{ST}$}&\quad&\multicolumn{3}{c}{$JM_{SSL}$}&\quad&\multicolumn{3}{c}{$JM_{SCN}$}\\ \cline{3-5} \cline{7-9} \cline{11-13} \cline{15-17}
		
		\quad&  & Estimate  & Lower & Upper &\quad& Estimate & Lower & Upper &\quad& Estimate  & Lower & Upper&\quad& Estimate & Lower & Upper  \\
		
		\hline
		\vspace{0.3cm}
		\textbf{For longitudinal} \\

		Intercept $(\beta_{11})$ & \quad &2.6170  & 2.4980 &2.7460& \quad &2.6430 & 2.5130 &2.7820&\quad &2.6470 & 2.5330&2.7610& \quad &2.6260 & 2.4920& 2.7400  \\
		
		Obs. Time ($\beta_{12}$)& \quad &-0.4322 & -0.5245 &-0.3381& \quad & -0.4345 & -0.5272 & -0.3274&\quad &-0.4505& -0.5330&-0.3633& \quad &-0.4496 & -0.5408 & -0.3579  \\
		
		Drug ($\beta_{13}$)& \quad & 0.0663 & -0.0535 & 0.2121& \quad &0.0644 & -0.0606 &0.1811&\quad &0.0620 & -0.0575&0.1753& \quad &0.0711 & -0.0563 & 0.1945  \\
		
		Gender ($\beta_{14}$)& \quad & -0.0082 & -0.1172 &0.0971& \quad &-0.0016 & -0.1266 &0.0969&\quad &-0.0118 & -0.1223&0.1088& \quad &0.0151 & -0.0937 & 0.1254  \\
		
		 Prev Opp Infection($\beta_{15}$)& \quad & -0.4340 & -0.5026 & -0.3567& \quad &-0.4264 & -0.4974 &-0.3381&\quad &-0.4365 & -0.5217&-0.3465& \quad &-0.4347 & -0.5154 & -0.3541  \\
		
		Intol to ZDV ($\beta_{16}$)& \quad &-0.0319 & -0.1172 & 0.0532& \quad &-0.0245 & -0.1111 &0.0589&\quad & -0.0213 & -0.1008&0.0490& \quad & -0.0248 & -0.1040 & 0.0510  \\
		\vspace{0.3cm}
		
		precision ($\tau_z$) & \quad &7.7740 & 6.9730 &8.6370& \quad &16.840 & 14.140 &20.110&\quad &27.310 & 22.950&31.960& \quad &16.680 & 14.280 & 19.300  \\
		
\vspace{0.3cm}
		\textbf{For both processes} \\
		
		Variance component ($\Omega_1$)& \quad &2.1540 & 1.8760 &2.5180& \quad &7.7130 & 2.2530 &15.120&\quad &6.4100 &2.8390& 15.280& \quad &16.500 & 2.0770 & 32.890  \\
		
		Covariance ($\Omega_2$)& \quad & 0.5008 &0.3072 &0.6990& \quad &0.5006 & 0.2979 &0.6929&\quad & 0.4976 & 0.2982 &0.7022 & \quad &0.5001 & 0.3087 & 0.6939 \\
		
		skewness parameter ($\lambda_1$)& \quad &1.0280 & 0.0818 &1.9550& \quad &1.2250 & 0.2336 & 1.9680&\quad &1.2990 & 0.3040&1.9670& \quad &1.0140 & 0.0723 & 1.9470  \\
		\vspace{0.1cm}
		skewness parameter ($\lambda_2$)& \quad &1.0010 & 0.0681 &1.9490& \quad &1.009 & 0.0627 &1.9520&\quad &0.9870 & 0.0634&1.9440& \quad &1.0160 & 0.0715 & 1.9390  \\
		
		\vspace{0.2cm}
		Cov between two processes ($\sigma^2_{cov}$)& \quad &0.2967 &0.0138 & 0.5826& \quad &0.3005 & 0.0147 &0.5874&\quad & 0.2989 & 0.0170&0.5830& \quad &0.3012 & 0.0129 & 0.5832  \\
		\vspace{0.2cm}
		
	\textbf{For time-to-event} \\

		intercept ($\beta_0$)& \quad &1.0000 & 0.8042 &1.2040& \quad &0.9981  & 0.7985 &1.1980&\quad & 0.9998 & 0.8003&1.1870 & \quad &1.0010 & 0.8026 & 1.2000  \\
		random intercept ($\nu_1$)& \quad & 0.9969 & 0.8026 &1.1900& \quad &0.9995 & 0.8027 &1.1950&\quad & 1.002 & 0.7991&1.1960& \quad & 0.9990 & 0.8164 & 1.1910  \\
		random slope ($\nu_2$)& \quad &0.9999 & 0.7966 &1.2050& \quad &0.9977 & 0.8059 &1.1950&\quad &0.9992 & 0.8019&1.1950& \quad & 1.0010 & 0.8066 & 1.1980  \\

		precision ($\tau_t$)  & \quad &0.9997 & 0.8131 &1.2130& \quad & 16.840 & 14.140 &20.110&\quad &1.0020& 0.8230&1.2100& \quad & 1.0000 & 0.8134 & 1.2030  \\
		\hline
		
	\end{tabular}
\caption{Data Study: Aids Data}\label{tab:data}
	\label{tab:aidsskew}
\end{table}

\end{document}